\documentclass[journal,a4paper,comsoc]{IEEEtran} 
\usepackage{amsmath,amsfonts}
\usepackage{algorithmic}
\usepackage{algorithm}
\usepackage{array}
\usepackage{textcomp}
\usepackage{stfloats}
\usepackage{url}
\usepackage{verbatim}
\usepackage{graphicx}
\usepackage{cite}
\usepackage{amssymb}
\usepackage{amsthm} 
\newtheorem{proposition}{Proposition}
\usepackage{booktabs}
\usepackage{multirow} 
\usepackage{tabularx}
\usepackage{makecell}
\usepackage{color,soul}
\usepackage{diagbox}
\usepackage{balance}

\DeclareRobustCommand{\bigO}{%
  \text{\usefont{OMS}{cmsy}{m}{n}O}%
}

\newtheorem{lemma}{Lemma}
\newtheorem{definition}{Definition}
\newtheorem{remarkex}{Remark}
\newenvironment{remark}
{\pushQED{\qed}\remarkex}
{\popQED\endremarkex}

\newcommand{\TitleFont}{\fontsize{25pt}{28pt}\selectfont}

\begin{document}

\title{\TitleFont Energy Efficiency Maximization for Integrated Sensing and  Communication in IoT C-RAN}

\author{Le Tung~Giang, Nguyen Xuan~Tung, Trinh Van~Chien, \\Chau~Yuen,~\IEEEmembership{Fellow,~IEEE}, and Won-Joo~Hwang,~\IEEEmembership{Senior Member,~IEEE}
\thanks{This work was supported by the National Research Foundation of Korea (NRF) grant funded by the Korea government (Ministry of Science and Information and Communication Technology (MSIT)) (RS-2024-00336962)
This work was supported by the Korea Agency for Infrastructure Technology Advancement (KAIA) grant funded by the Ministry of Land Infrastructure and Transport (No. RS-2023-00256816).
This work was supported by Institute of Information \& communications Technology Planning \& Evaluation (IITP) under the Artificial Intelligence Convergence Innovation Human Resources Development (IITP-2025-RS-2023-00254177) grant funded by the Korea government (MSIT).
This work was partly supported by the Institute of Information \& Communications Technology Planning \& Evaluation (IITP)-ITRC (Information Technology Research Center) grant funded by the Korea government (MSIT) (IITP-2025-RS-2023-00260098, contribution rate: 25\%).
\textit{(Corresponding author: Won-Joo~Hwang.)}}
\thanks{Le Tung Giang is with the Department of Information Convergence Engineering, Pusan National University, Busan 46241, Republic of Korea (email: giang.lt2399144@pusan.ac.kr).}
\thanks{Nguyen Xuan Tung is with the Faculty of Interdisciplinary Digital Technology, PHENIKAA University, Yen Nghia, Ha Dong, Hanoi 12116, Vietnam (email: tung.nguyenxuan@phenikaa-uni.edu.vn). }
\thanks{Trinh Van Chien is with the School of Information and Communications Technology, Hanoi University of Science and Technology, Hanoi 100000, Vietnam (email: chientv@soict.hust.edu.vn).}
\thanks{Chau Yuen is with the School of Electrical and Electronic Engineering, Nanyang Technological University, Singapore 639798 (email: chau.yuen@ntu.edu.sg).}
\thanks{Won-Joo Hwang is with the School of Computer Science and Engineering, Center for Artificial Intelligence Research, Pusan National University, Busan 46241, South Korea (email: wjhwang@pusan.ac.kr).}
}


\maketitle

\begin{abstract}
Integrated sensing and communication (ISAC) with cloud radio access networks (C-RAN) unifies two foundational technologies for the Internet of Things (IoT), enabling both high-data-rate transmission and accurate environmental awareness. This paper investigates an uplink C-RAN system where multiple remote radio units (RRUs) jointly serve user equipment devices (UEs) and sense a target. We demonstrate that only a subset of RRUs is typically needed to meet sensing requirements, while energy is potentially conserved by deactivating redundant sensing RRUs. Motivated by this, we aim to improve the system's energy efficiency (EE) by jointly optimizing RRU activation, user association, and power allocation, while ensuring localization accuracy through the Cramér-Rao lower bound (CRLB). To address this problem, we propose a model-based iterative algorithm that integrates optimal user association, decision tree-based selection of sensing RRUs, and fixed-point power allocation. This algorithm is further used to train a graph neural network (GNN) framework, ISAC-GNN, to overcome scalability and computational complexity limitations. Our theoretical analysis ensures performance guarantees, while a semi-supervised training strategy enhances stability and generalization. Simulation results demonstrate that ISAC-GNN reduces inference time by $\mathbf{67\%}$ relative to the model-based approach, enabling real-time implementation for ISAC system resource management. In addition, ISAC-GNN allows scalable deployment across various system scenarios without retraining the model.
\end{abstract}

\begin{IEEEkeywords}
C-RAN, graph neural network, integrated sensing and communication, Internet of Things,  multi-static sensing.
\end{IEEEkeywords}

\vspace{-0.9cm}
\section{Introduction}

\IEEEPARstart{T}{he coexistence} of communication and sensing is indispensable for meeting the rising demands for high-capacity transmission and accurate environmental awareness in Internet of Things (IoT) systems \cite{2024_IOTJ_Lu_IsacSurvey}. 
Against this backdrop, \textcolor{black}{integrated sensing and communication} (ISAC) has emerged as a promising technique to support various IoT applications, including smart cities \cite{2024_IOTJ_Li_TrainISAC}, industrial IoT \cite{2025_IOTJ_Jing_ISAC_Edge}, and human tracking \cite{2025_IOTJ_Wang_HumanTrack}.
The integration of communication and sensing has been extensively studied, where early designs often separated these functions in time, frequency, or code domains \cite{2013_IET_Han_Sensing}, thereby avoiding mutual interference but sacrificing spectral efficiency. To overcome this limitation, passive sensing has emerged as a compelling alternative \cite{2024_TCOM_Jiang_PassiveSensing}, leveraging received communication signals for sensing tasks with minimal modifications to the transmit waveform and only minor adjustments to the receiver design. This approach enables seamless ISAC deployment while making full use of existing IoT infrastructure \cite{2025_IOTJ_Lee_PimISAC, 2025_IOTJ_OfdmIsac, 2024_IOTJ_Jiang_NomaIsac, 2025_IOTJ_Du_dlMimoIsac, 2024_COML_ISAC_UAV}.

ISAC was initially explored in single-cell IoT scenarios with monostatic sensing, where a single IoT access point (AP) performs both communication and sensing tasks \cite{2025_IOTJ_Zhou_StarRisIsac}. While suitable for basic deployments, this setup offers limited sensing coverage and is highly susceptible to blockage, making it inadequate for dynamic and complex IoT environments such as smart cities or industrial automation \cite{2025_IOTJ_Niu_IsacSurvey}.
To overcome these limitations, many studies have explored bistatic sensing, where the transmitters and receivers are non-co-located, improving coverage and robustness \cite{2024_TVT_Nataraja_BiISAC, 2024_TVT_Hu_BiISAC}.
Building on this, multistatic sensing further enhances performance by leveraging multiple non-co-located transmitters and receivers. This configuration enables coordinated sensing and joint signal processing across distributed IoT APs, significantly improving detection range, sensing resolution, and environmental awareness \cite{2025_TVT_Li_MultiBsUAV, 2024_WCMag_Meng_CoopISAC}.

While ISAC can be deployed on various distributed wireless architectures, the \textcolor{black}{cloud radio access network} (C-RAN) stands out as a particularly promising platform for large-scale, cooperative ISAC systems. By combining spatially distributed \textcolor{black}{radio remote units} (RRUs) with fully centralized baseband processing, C-RAN inherently supports cooperative multistatic ISAC operations.
Moreover, the cloud infrastructure in C-RAN offers enhanced security through robust encryption, access control, and centralized data handling, reducing privacy risks associated with decentralized data processing.
Despite these advantages, implementing cooperative ISAC in C-RAN presents new challenges, including synchronization between transmitters and receivers \cite{2024_TSP_Larsson_Sync}, signaling load balance among RRUs \cite{2024_WCL_Zou_SignalLoad}, and
coordinated power allocation. While the first issue is handled by implementing a fully synchronized infrastructure \cite{2021_Book_Demir}, the latter two resource management problems remain open and demand advanced, scalable solutions due to the system's escalating complexity.

In ISAC-enabled C-RAN systems, resource management often focuses on maximizing system performance metrics such as achievable communication rate or sensing precision. However, with the growing demand for data traffic, energy consumption in wireless networks has risen sharply \cite{2023_TWC_Chen_EE}, and this becomes even more critical in joint communication and sensing scenarios due to additional sensing-related processing tasks. Therefore, many works have investigated resource management with a focus on energy efficiency (EE) in ISAC C-RAN systems \cite{2024_TWC_Behdad_ISACcfmMIMO, 2024_TWC_Cheng_ISAC_distributedMIMO}.
In \cite{2024_TWC_Behdad_ISACcfmMIMO}, the authors focus on power allocation to maximize the sensing signal-to-interference-plus-noise (SINR) ratio. Meanwhile, the study \cite{2024_TWC_Cheng_ISAC_distributedMIMO} considers user association to manage interference and maximize the sensing minimum detection probability.
Even though traditional optimization algorithms can achieve near-optimal solutions for resource management, they often suffer from high computational complexity \cite{2024_Tung_SOICT}. This limits their practicality for real-time implementation in large-scale ISAC systems \cite{TVT_ISAC_Huang_Jul_2022}. To address this issue, deep neural networks (DNNs) have emerged as flexible and adaptive alternatives \cite{2025_IOTJ_Du_dlMimoIsac}. Among them, graph neural networks (GNNs) are particularly promising due to their ability to capture structural relationships between network elements \cite{2024_Tung_GNNSurvey}. By modeling the wireless network as a graph with nodes and edges, GNNs can generate predictions independently of the system size, making them well-suited for complex multi-transmitter systems \cite{2024_TVT_Tung_GNN_cfmMIMO} and dense large-scale wireless networks \cite{2024_SensJ_Giang_HGNN}. 

Additionally, most existing ISAC studies primarily focus on the downlink (DL) scenario, with limited attention given to uplink (UL) ISAC. Nonetheless, the consideration of the UL is crucial for enabling full-duplex ISAC designs. A recent study \cite{2024_VTC_ISAC_UL} investigates power allocation and user scheduling in a UL cell-free ISAC system, demonstrating the feasibility of simultaneous communication and sensing using uplink signals.
However, prior works typically assume that all RRUs are active during the sensing process, leading to redundant power consumption and service overhead. Given that the sensing operation incurs significant power costs \cite{Boora2024}, activating only a necessary subset of RRUs for sensing is a more efficient strategy, effectively reducing the overall system power consumption. This observation highlights the importance of incorporating RRU sensing activation into the resource management framework to further improve EE in ISAC systems.

\begin{table}[t]
    \renewcommand{\arraystretch}{1.5}
    \centering
    \caption{Literature Overview on Multi-transmitter ISAC Networks.}
    \resizebox{\columnwidth}{!}{
    \begin{tabular}{ccccc}
        \hline
        \makecell{Comm.\\mode}   & \makecell{Power\\allocation}     & \makecell{User\\association}  &\makecell{RRU\\activation}  &Reference \\
        \hline
        \multirow{2}{*}{DL}       & $\boldsymbol{\checkmark}$      &-       &-     
        &\cite{2025_IOTJ_Zhang_THzISAC, 2025_IOTJ_Fan_cfMIMOISAC, 2024_TWC_Behdad_ISACcfmMIMO,2025_TVT_Tung_HetGNN, 2024_TCOM_Zou_EE_ISAC, 2024_IOTJ_ISAC_Power, 2024_TWC_ISAC_Power, 2025_TVT_Shu_IsacJam }\\
        
                \cline{2-5}
                &$\boldsymbol{\checkmark}$       &$\boldsymbol{\checkmark}$       &-   
        & \cite{2024_TWC_ISAC_PowerUser, 2024_TWC_Cheng_ISAC_distributedMIMO, 2025_TWC_Zhang_UAV_Beam}\\
        \hline
        \multirow{2}{*}{UL}        &$\boldsymbol{\checkmark}$       &$\boldsymbol{\checkmark}$       &-    
        & \cite{2024_VTC_ISAC_UL}\\
                \cline{2-5}
                &$\boldsymbol{\checkmark}$       &$\boldsymbol{\checkmark}$       &$\boldsymbol{\checkmark}$   
        & \textbf{This Paper}\\
        \hline
    \end{tabular}
    }
    \label{tab:prev_works}
    \vspace{-0.5cm}
\end{table}

\textcolor{black}{In this context, this paper considers the EE in an uplink C-RAN integrated with passive sensing. 
Since user equipment devices (UEs) have limited power, we aim to maximize the total EE over all network devices, including RRUs and UEs.
And to extensively improve this, we incorporate RRU activation alongside power allocation and user association.} 
Table~\ref{tab:prev_works} categorizes existing studies on resource management in ISAC with distributed multi-transmitter systems, based on their utility functions and optimization variables.
To the best of our knowledge, this is the first work to consider sensing receiver power consumption in an uplink C-RAN employing multistatic passive sensing.
We jointly optimize power allocation, user association, and RRU activation to maximize the overall EE across all RRUs and UEs, while meeting the sensing accuracy constraint.
In addition to the proposed model-based iterative algorithm, we introduce a GNN-based framework, termed Integrated Sensing and Communication GNN (ISAC-GNN), which jointly learns the optimal resource management policies using a semi-supervised training strategy.
The main contributions of this paper are summarized as follows

\begin{itemize} 
\item We formulate an EE maximization problem for ISAC in C-RAN. In particular, three resource management variables are jointly considered: (i) power allocation, (ii) user association, and (iii) RRU activation to enhance system-wide EE.  \textcolor{black}{In addition to resource constraints, the problem ensures reliable passive sensing by incorporating the Cramér-Rao lower bound (CRLB), a widely employed, analytically tractable metric for sensing accuracy in prior ISAC and radar system design works.} Due to the non-convex nature of both the objective and CRLB constraints, the problem cannot be directly solved and requires advanced optimization techniques.

\item For the model-based approach, we propose an iterative algorithm to solve the formulated EE maximization problem by decomposing and solving the sub-problems of user association, RRU activation, and power allocation. For the user association, we construct an algorithm to provide the optimal association between UEs and RRUs. For the RRU activation, we introduce a decision tree-based approach to determine the solution subset of active RRUs. Additionally, a fixed-point algorithm is proposed to efficiently compute the optimal power allocation for all UEs, ensuring convergence and improved overall EE.

\item To address the scalability and real-time implementation, we propose a GNN framework, namely ISAC-GNN, jointly solving the aforementioned EE maximization problem. The ISAC-GNN leverages a heterogeneous graph representation of the ISAC C-RAN system, capturing the complex interactions between UEs and RRUs. It inherits the scalability and generalization properties of GNNs while significantly reducing the time complexity compared to the model-based approach.

\item We provide a theoretical analysis of the ISAC-GNN architecture, guaranteeing performance and demonstrating its effectiveness in solving the EE maximization problem. This analysis ensures the reliability and robustness of the proposed learning-based GNN framework.

\item To improve the training efficiency and performance of the proposed ISAC-GNN, we introduce a semi-supervised training strategy. First, this approach initializes the parameters using supervised training, using guidance from the proposed model-based approach for a stable starting point. Then, the process transitions smoothly to unsupervised training to better maximize the reward of the desired problem. This transition between the two training phases ensures both stability during initialization and efficiency in optimizing the objective.
\end{itemize}

The remainder of this paper is structured as follows: Section \ref{sec:SystemModel} presents the integrated C-RAN and passive sensing system model, incorporating the CRLB sensing accuracy requirement. The EE maximization problem is formulated under multiple resource management policies. Section \ref{sec:OptimiaztionApproach} introduces a model-based algorithm to solve this problem, while Section \ref{sec:ISAC-GNN} proposes a scalable and real-time applicable learning-based solution. Theoretical performance analysis is provided in Section \ref{sec:Theoretical}. Numerical results are discussed in Section \ref{sec:Sim}, and conclusions are drawn in Section \ref{sec:conclusion}.

\emph{Notation:} We use boldface lowercase letters and boldface uppercase letters to denote vectors and matrices, respectively. Let $\mathbf{a}^T$ denote the transpose of $\mathbf{a}$. The component-wise inequality between vectors $\mathbf{a}$ and $\mathbf{b}$ is written as $\mathbf{a}\succeq\mathbf{b}$. 
We denote the circularly symmetric complex Gaussian distribution with zero mean and variance $\sigma^2$ by $\mathcal{CN}(0,\sigma^2)$.

\section{System Model and Problem Formulation}\label{sec:SystemModel}
\begin{figure}
    \includegraphics[width=\columnwidth]{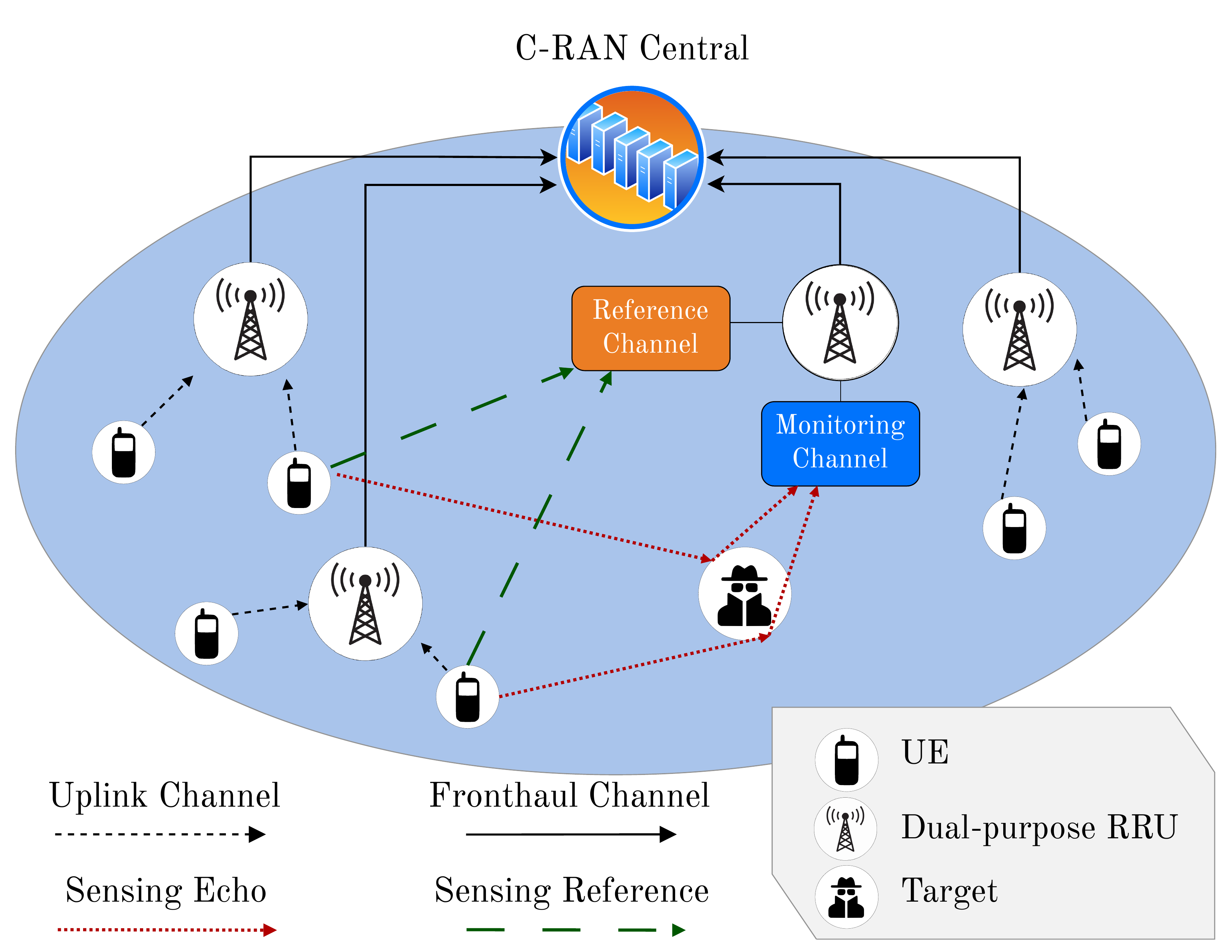}
    
    \caption{\textcolor{black}{The ISAC C-RAN system architecture.}}
    \label{SystemModelFigure}
\end{figure}

\textcolor{black}{Consider a distributed uplink C-RAN system consisting of a set of $N$ UEs $\mathcal{N}~\triangleq~\{1,\dots,N\}$, a set of $K$ dual-purpose RRUs $\mathcal{K}~\triangleq~\{1,\dots,K\}$ and a single target \footnote{This framework is potentially extended to multi-target scenarios. However, to maintain clarity and focus in this work, we leave such extensions for future study.} as illustrated in Fig.~\ref{SystemModelFigure}}.
In this work, we adopt a partially centralized C-RAN architecture to enhance system scalability and reduce the centralized computation bottleneck. Specifically, each UE’s uplink payload is decoded locally at a single serving RRU, while the centralized cloud performs network-level coordination and multi-static sensing fusion \cite{2025_IOTJ_PartialCRAN, 2023_TWC_PartialCRAN}.
During each time-frequency block, RRUs receive uplink signals from UEs, decode the signals of their associated UEs using local observations, and forward the decoded information to a centralized cloud processor.
These uplink signals are reused for passive sensing \cite{Passive_Sensing_Zhang_2024}, where each RRU employs a two-channel scheme to capture both the target's echo and the direct uplink signals. By comparing them, key target parameters such as location, velocity, and angle of arrival (AoA) are estimated \cite{Malanowski2012}. 
Accordingly, each RRU forwards (i) the decoded uplink payload of its associated UEs and (ii) sensing-related observations extracted from the two-channel receiver to the centralized cloud via ideal, fully synchronized fronthaul links \cite{chien2020}.
Key notations are summarized in Table~\ref{tab:notations}.

\begin{table}[t]
    \centering
    \caption{Table of Abbreviations and Notations}
    \setlength{\tabcolsep}{1em} 
    \renewcommand{\arraystretch}{1.2}
    \begin{tabular}{|c|p{0.75\columnwidth}|}
        \hline
        \textbf{Notation} & \textbf{Description} \\ \hline
        $\mathcal{N}$ & Set of user equipment devices (UEs). \\ \hline
        $\mathcal{K}$ & Set of remote radio units (RRUs). \\ \hline
        $P_{n}^{\textrm{max}}$ & Maximum transmit power of UEs.\\ \hline
        $P_{\textrm{cir}}$ & Circuit power consumption of UEs.\\ \hline
        $P_{\textrm{sens}}$ & Sensing power consumption of RRUs.\\ \hline
        $[\hat{x}_{n}, \hat{y}_{n}]$ & Coordinates of the $n$-th UE. \\ \hline
        $[\tilde{x}_{k}, \tilde{y}_{k}]$ & Coordinates of the $k$-th RRU. \\ \hline
        $[x, y]$ & Coordinates of the single target. \\ \hline
        $\hat{R}_{n}$ & The distance between the $n$-th UE and the single target. \\ \hline
        $\tilde{R}_{k}$ & The distance between the $k$-th RRU and the single target. \\ \hline
        $g_{kn}$ & Channel gain between the $k$-th RRU and the $n$-th UE. \\ \hline
        \multirow{1.5}{*}{$\rho_{kn}$} & The radar cross section (RCS) between the $n$-th UE and the $k$-th RRU with respect to the target. \\ \hline
        $\boldsymbol{\Psi}$ & RRU selection variable. \\ \hline
        $\mathbf{p}$ & Transmit power of the UEs. \\ \hline
        $\boldsymbol{\tau}$ & RRU sensing activation variable. \\ \hline
        $\gamma_{n}$ & SINR of the $n$-th UE. \\ \hline
        $\sigma^2$ & Variance of additive white Gaussian noise (AWGN). \\ \hline
        $\kappa$    & Maximum acceptable localization error. \\ \hline
    \end{tabular}
    \label{tab:notations}
    \vspace{-0.3cm}
\end{table}

\subsection{Communication Subsystem}

During the uplink transmission, the $n$-th UE sends signal to its selected RRU with the allocated power $0\leq P_n\leq P_{n}^{\textrm{max}}$. The received signal at the $k$-th RRU is given by

\begin{equation}\label{ReceiveSN}
    y_{k}=\sum_{n\in \mathcal{N}} g_{kn} \sqrt{P_{n}} s_{n} + n_{k},
\end{equation}
where $g_{kn}$ denotes the channel gain between the $n$-th UE and the $k$-th RRU, and $s_n$ with $\mathrm{E\{\vert s_{n}\vert ^2\}} = 1$ is the zero-mean signal from the $n$-th UE, and the last term $n_{k}\sim \mathcal{CN}(0,\sigma^2)$ is the additive white Gaussian noise (AWGN) at the $k$-th RRU, \textcolor{black}{with the variance $\sigma^2=B N_0$}. Here, $N_0$ and $B$ denote the noise power spectral density and the bandwidth of each UE, respectively. 
From the perspective of the $n$-th UE, the received signal at its selected $k$-th RRU is given by
\begin{align}
    y_{k}= \underbrace{\underset{\phantom{\sum_{n_{A}}}}{g_{kn} \sqrt{P_{n}} s_{n}}}_{\mathsf{DS}} + \underbrace{\sum_{n^{\prime}\in \{\mathcal{N}\setminus n\}} g_{kn^{\prime}} \sqrt{P_{n^{\prime}}} s_{n^{\prime}}}_{\mathsf{MI}} + \underbrace{\underset{\phantom{\sum_{n_{A}}}}{n_{k}}}_{\mathsf{Noise}},
\end{align}
where $\mathsf{DS}$ denotes the desired signal transmitted from the $n$-th UE, and $\mathsf{MI}$ represents the mutual interference caused by all other UEs at the selected RRU.
The total achievable rate of all UEs is then computed as
\begin{equation}\label{RateCalculation}
    \begin{aligned}
        R_{\textrm{tot}}=\sum_{n\in \mathcal{N}}&R_n=\sum_{n\in \mathcal{N}}B\log_2 \left(1+\gamma_{n}\right),
    \end{aligned}
\end{equation}
where $\gamma_{n}$ is the signal-to-interference-plus-noise (SINR) ratio for the $n$-th UE, which is computed as
\begin{align}\label{SINR}
    \gamma_{n} &= \frac{\vert\mathsf{DS}\vert^{2}}{\vert\mathsf{MI}\vert^{2} + \sigma^2} =\frac{\sum_{k\in\mathcal{K}}\vert\mu_{kn}g_{kn} \vert^{2}P_{n}}{\sum_{n^{\prime}\in \{\mathcal{N}\setminus n\}}\sum_{k\in\mathcal{K}}\vert\mu_{kn'}g_{kn} \vert^{2}P_{n'} +\sigma^2} \nonumber\\ 
    &=\frac{ \vert\boldsymbol{\mu}_{n}^T \boldsymbol{g}_n\vert^2 P_{n}}{ \sum_{n'\in \mathcal{N}\setminus \{n\}}\vert \boldsymbol{\mu}_n^T \boldsymbol{g}_{n'}\vert ^2P_{n'}+\sigma^2},
\end{align} 
where we define $\boldsymbol{g}_{n}=[g_{1n},...,g_{Kn}]^T$, and $\boldsymbol{\mu}_n=[\mu_{1n},...,\mu_{Kn}]^T \in \{0,1\}^{K}$ represents the RRU selection of the $n$-th UE. 
In particular, if the uplink data of the $n$-th UE is decoded at the $k$-th RRU then $\mu_{kn}=1$, and otherwise, $\mu_{kn}=0$  means the signal from the $n$-th UE is not decoded at the $k$-th RRU.  
The signal of each UE will be decoded at one RRU only, and thus $\sum_{k\in \mathcal{K}}\mu_{kn}=1,\hspace{0.1cm} \forall n\in \mathcal{N}$.
The decoded data are then forwarded to the cloud for further centralized processing.

\subsection{Radar Sensing Subsystem}

In the sensing context, this ISAC system detects target devices and estimates their locations using a distributed multistatic sensing setup, where transmitters and receivers are not co-located.
\textcolor{black}{We denote the locations} of the target, UEs, and RRUs in a two-dimensional (2-D) coordinate system as $(x,y)$, $(\hat{x}_{n},\hat{y}_{n})$, and $(\tilde{x}_{k},\tilde{y}_{k})$, respectively, where $n\in\mathcal{N}$, and $k\in\mathcal{K}$. The propagation delay from the $n$-th UE to the target and reflected to the $k$-th RRU is expressed as $\Delta t_{kn} =  (\hat{R}_{n} +  \tilde{R}_{k})/c$,
where $c=3\times 10^{8}\mbox{ [m/s]}$ is the \textcolor{black}{signal propagation speed}, $\hat{R}_{n} = \sqrt{(\hat{x}_{n}-x)^2 + (\hat{y}_{n}-y)^2}$ is distance between the target and the $n$-th UE, $\tilde{R}_{k} = \sqrt{(\tilde{x}_{k}-x)^2 + (\tilde{y}_{k}-y)^2}$ is the distance between the target and the $k$-th RRU. 
In practical deployments, the sensing receiver collects a superposition of the desired target echo, a strong direct-path UE-RRU component, and multipath reflections from environmental scatterers. 
Nevertheless, under the widely adopted two-channel passive sensing receiver architecture, each RRU obtains a reference copy of the uplink signal, which enables reference-assisted direct-path and clutter suppression.
In particular, the reference copy is used to coherently estimate the dominant direct-path component via time-delay and frequency alignment, and the estimated component is subtracted from the monitoring channel~\cite{2022_Sensors_DPCancel}.
Subsequently, residual reflections from stationary or slowly varying environmental scatterers are mitigated using standard clutter suppression techniques, such as adaptive filtering, Doppler-domain, and space-time domain processing methods~\cite{2025_IOTJ_clutter}. 
Accordingly, the post-cancellation residual signal at the $k$-th RRU is expressed as
\begin{align}\label{eq:sensing_re}
    r_{k}(t) = \sum_{n = 1}^{N} \rho_{kn} \sqrt{P_n}s_{n}(t- \Delta t_{kn}) + w_{k}(t),
\end{align}
where the dominant component corresponds to the target echo, $w_{k}(t)\sim \mathcal{CN}(0,\sigma^2)$ represents effective noise after imperfect suppression at the monitoring channel; $\rho_{kn}$ denotes the radar cross-section (RCS) for the path from the $n$-th UE to the target and back to the $k$-th RRU, which we assume to be independent across distinct $(k,n)$ pair.
As outlined in \cite{Sensor_SPM_2007}, radar systems operate in two primary modes: (i) \textit{searching mode}, where no prior CSI is available and waveforms scan the angular domain for detection; and (ii) \textit{tracking mode}, where previously estimated target states refine position estimates over time.
To focus on resource allocation, we consider the ISAC system in tracking mode, where the target RCS ${\rho_{kn}}$ and location $(x,y)$ are known and continuously estimated through each time step as $ \theta = [x,y, \pmb{\rho}]$, where $\pmb{\rho} = \{\rho_{kn} \vert 1 \leq n \leq N, 1 \leq k \leq K \}$. This two-phase sensing process remains sufficient under dynamic target tracking for practical time-varying deployments.


\begin{remark}
    In the considered two-phase sensing system, low SNR and interference may bias the estimator, causing actual error to exceed the CRLB in the first phase. However, in the second phase, CRLB is tighter when tracking the target \cite{Godrich2011, Godrich2010, 2022_JSAC_Liu_ISAC}. Furthermore, alternative sensing metrics such as end-to-end MSE or posterior CRLB often lack such closed-form dependence or require embedding a full estimator in the optimization loop \cite{Sensor_SPM_2007}. Meanwhile, the CRLB offers a closed-form, analytically tractable expression that directly depends on key system parameters. Accordingly, using the CRLB is both technically justified and readily extendable to more sophisticated system models. 
\end{remark}

Following \cite{Godrich2010}, for the vector parameter $\theta$, the closed-form CRLB matrix $\textbf{C}_{x,y}$ is defined as the submatrix of the inverse of the Fisher information matrix (FIM) as follows
\begin{align}\label{eq:crlb}
    \textbf{C}_{x,y} (\mathbf{p}, \boldsymbol{\tau}) = \left\{ \sum_{n \in \mathcal{N}} P_{n} 
    \begin{bmatrix}
    q_{a_n}(\boldsymbol{\tau}) & q_{c_n}(\boldsymbol{\tau}) \\
    q_{c_n}(\boldsymbol{\tau}) & q_{b_n}(\boldsymbol{\tau}) 
    \end{bmatrix} \right\}^{-1},
\end{align}
where the definitions hold $\mathbf{p}=[P_1,...,P_N]^T$, and the RRU sensing activation $\boldsymbol{\tau} = [\tau_{1}, \dots, \tau_{K}]$ is a binary vector representing the participation status of all RRUs in the passive sensing process. 
Each element $\tau_{k} \in \{0,1\}$ corresponds to the state of the $k$-th RRU, where $\tau_{k} = 1$ indicates that the RRU is active and participates in the sensing process, while $\tau_{k} = 0$ indicates that the RRU is inactive and does not contribute to sensing. The elements $q_{a_n}(\boldsymbol{\tau}), q_{b_n}(\boldsymbol{\tau}), \text{ and } q_{c_n}(\boldsymbol{\tau})$ respectively are defined as

\begin{subequations}
\begin{align}
    q_{a_n}(\boldsymbol{\tau}) &= \zeta \sum_{k\in \mathcal{K}} \tau_{k} \vert \rho_{kn}\vert ^2 \left( \frac{\hat{x}_{n} - x}{\hat{R}_{n}} + \frac{\tilde{x}_{k} - x}{\tilde{R}_{k}} \right)^2, \\
    q_{b_n}(\boldsymbol{\tau}) &= \zeta \sum_{k\in \mathcal{K}} \tau_{k} \vert \rho_{kn}\vert ^2 \left( \frac{\hat{y}_{n} - y}{\hat{R}_{n}} + \frac{\tilde{y}_{k} - y}{\tilde{R}_{k}} \right)^2, \\
    q_{c_n}(\boldsymbol{\tau}) &= \zeta \sum_{k\in \mathcal{K}} \tau_{k} \vert \rho_{kn}\vert ^2 \left( \frac{\hat{x}_{n} - x}{\hat{R}_{n}} + \frac{\tilde{x}_{k} - x}{\tilde{R}_{k}} \right) \nonumber\\ 
            & \qquad \times \left( \frac{\hat{y}_{n} - y}{\hat{R}_{n}} + \frac{\tilde{y}_{k} - y}{\tilde{R}_{k}} \right) \quad, \forall n \in \mathcal{N},
\end{align}
\end{subequations}
with $\zeta = \frac{8\pi^2 B^2}{\sigma_{S}^{2} c^2}$. 
The sum of CRLBs for estimating the location of the target is expressed as
\begin{align}
    \sigma_{x,y}^2(\mathbf{p}, \boldsymbol{\tau}) =  \frac{\mathbf{b}^T \mathbf{p}}{\mathbf{p}^T \mathbf{A} \mathbf{p}},
\end{align}
where $\mathbf{b} = \textbf{q}_{a} + \textbf{q}_{b}$, $\mathbf{A} = \textbf{q}_{a}\textbf{q}_{b}^T - \textbf{q}_{c}\textbf{q}_{c}^T$,
with $\textbf{q}_{a} = [q_{a_1},\dots,q_{a_N}]^T, \textbf{q}_{b} = [q_{b_1},\dots,q_{b_N}]^T, \text{ and } \textbf{q}_{c} = [q_{c_1},\dots,q_{c_N}]^T$. 
\textcolor{black}{The estimations of $\theta$ are updated in real time based on the previous time step estimation.}

\begin{remark}
    In \cite{Godrich2010}, the closed-form CRLB is derived under the assumption that the sensing waveforms are orthogonal, which eliminates cross-correlation terms in FIM, yielding the minimum achievable lower bound on the localization error \cite[Sec.~III.C]{Godrich2010}. By contrast, the considered uplink C-RAN sensing subsystem operates with non-orthogonal UE uplink transmission. Nevertheless, since the UE-specific pilot components are known at the RRUs, pilot-aided multiuser separation and interference cancellation can be applied to suppress the cross-user coupling terms, resulting in a residual observation that is approximately separable across UEs. Such pilot-aided cancellation therefore eliminates the corresponding cross-correlation terms in the FIM.
    Accordingly, this facilitates using the closed-form CRLB expressed in \eqref{eq:crlb} as a conservative design constraint for the sensing subsystem, even when non-orthogonal uplink transmissions are employed.
\end{remark}

\subsection{Problem Formulation}
\textcolor{black}{The total power consumed by the $n$-th UE is expressed as} \cite{Daehan2013} $P^{\textrm{con,UE}}_{n}=P_{\textrm{cir}} + P_{n}$, where $P_{\textrm{cir}}$ is the circuit power consumption at each UE for amplification and cooling purposes.
And the \textcolor{black}{power consumed by the $k$-th RRU is expressed as \cite{Boora2024} $P^{\textrm{con,RRU}}_{k} = \tau_{k} P_{\textrm{sens}}$, where $P_{\textrm{sens}}$ is the power dedicated to sensing at each RRU.}
We then compute the total power consumption of all UEs and RRUs as
\begin{align}\label{TotalP}
\begin{aligned}
    P_{\textrm{tot}}&= P^{\textrm{con,UE}} + P^{\textrm{con,RRU}} = \sum_{n\in \mathcal{N}}P^{\textrm{con,UE}}_n + \sum_{k\in \mathcal{K}}P^{\textrm{con,RRU}}_{k} \\
    &= \sum_{n\in\mathcal{N}}(P_{\textrm{cir}}+P_{n}) + \sum_{k\in\mathcal{K}}\tau_{k}P_{\textrm{sens}}.
\end{aligned}
\end{align}
The EE of the entire network is defined as
\begin{equation}\label{EEfirst}
    \eta(\boldsymbol{\Psi}, \boldsymbol{\tau}, \mathbf{p}) = \frac{R_{\textrm{tot}}}{P_{\textrm{tot}}},
\end{equation}
where $\boldsymbol{\Psi}=[\boldsymbol{\mu}_1,\dots,\boldsymbol{\mu}_N]$. In this work, the objective is to maximize the above EE over all devices, while both satisfying all system resource limitations and ensuring the desired localization sensing performance. Consider the maximum acceptable localization error $\kappa$.
The EE maximization problem of the ISAC C-RAN system is formulated as 
\begin{subequations}\label{Problemv1}
    \begin{align}
        && \underset{\substack{ \boldsymbol{\Psi}, \mathbf{p}, \boldsymbol{\tau} }}{\mathrm{maximize} } & \quad \eta(\boldsymbol{\Psi},\boldsymbol{\tau},\mathbf{p}) \\
        &&\mbox{s.t.}\hspace{0.15cm}&\quad 0 \leq P_{n} \leq P^{\textrm{max}}_{n} \quad, \forall n\in \mathcal{N},\label{V1b} \\
        &&&\quad \sum_{k\in\mathcal{K}}\tau_{k} \geq 1 \quad, \tau_{k}\in\{0,1\}, \forall k\in\mathcal{K}, \label{V1RRU}\\
        &&&\quad \sum_{k\in \mathcal{K}}\mu_{kn}=1\quad, \forall n\in \mathcal{N}, \label{V1c}\\
        &&&\quad  \sigma_{x,y}^2(\mathbf{p}, \boldsymbol{\tau})  \leq \kappa ,\label{V1d}
    \end{align}
\end{subequations} 
where constraint \eqref{V1b} is the transmit power limitation of each UE; constraint \eqref{V1RRU} represents the participation of RRU in the sensing process, guaranteeing at least one RRU is activated. \textcolor{black}{Meanwhile, constraint \eqref{V1c} indicates that the signal of each UE is decoded at only one RRU. 
Constraint \eqref{V1d} guarantees the target sensing accuracy.}
\textcolor{black}{Motivated by the CRLB approximation in \cite{Godrich2011}, the CRLB constraint \eqref{V1d} is first rewritten as $\mathbf{b}^T \mathbf{p} - \kappa \mathbf{p}^T \mathbf{A} \mathbf{p} \leq 0$, which is then relaxed to $\mathbf{b} - \kappa \mathbf{A} \mathbf{p} \preceq 0$.} \textcolor{black}{Accordingly, problem \eqref{Problemv1} is relaxed to the following problem}
\begin{subequations}\label{Problemv2}
    \begin{align}
        \underset{\substack{ \boldsymbol{\Psi}, \mathbf{p}, \boldsymbol{\tau} }}{\mathrm{maximize} } & \quad \eta(\boldsymbol{\Psi},\boldsymbol{\tau},\mathbf{p}) \label{P2:obj}\\
        \mbox{s.t.}\quad&\quad \eqref{V1b}, \eqref{V1RRU}, \eqref{V1c},  \nonumber \\
        &\quad  \mathbf{b} - \kappa \mathbf{A} \mathbf{p} \preceq 0 .\label{P2:sense}
    \end{align}
\end{subequations} 
However, this problem is still challenging to solve due to the non-convexity of the objective function, and an advanced algorithm design is needed. 

\begin{remark}
In the multi-target setting, the system model is naturally extended from the single-target case: instead of one reflected echo, multiple target echoes superpose at the receiver. This leads to a block-structured FIM, where each diagonal block captures per-target parameters while the off-diagonal blocks account for cross-target coupling effects. Such coupling reduces Fisher Information and adversely  tightens the power allocation constraints. Importantly, this system-level extension does not require new modeling assumptions, allowing the system to simply generalize from the single-target formulation to multiple targets.
\end{remark}

\section{Proposed Model-Based Approach}\label{sec:OptimiaztionApproach}
In this section, we address the optimization problem in \eqref{Problemv2}. The EE maximization problem is a non-convex mixed-integer program, making it challenging to obtain an optimal solution. To tackle this, we decompose \eqref{Problemv2} into three sub-problems: $(i)$ user association, $(ii)$ RRU sensing activation, and $(iii)$ power allocation. We provide solutions for each sub-problem and propose an Alternative Algorithm that sequentially optimizes them to efficiently solve the EE maximization problem.

\subsection{Initialization of Power Allocation $\normalfont{\mathbf{p}}$ and RRU Activation $\boldsymbol{\tau}$}

The user association problem is addressed first, requiring an initial power 
allocation and RRU activation setup. Initially, all UEs transmit at their 
maximum power, i.e., $P_n = P_n^{\textrm{max}}, \forall n \in \mathcal{N}.$
Similarly, all RRUs are assumed to be active in the sensing process by default, 
setting $\tau_k = 1, \forall k \in \mathcal{K}$. 

\subsection{User Association}\label{method:user_asso}
\begin{algorithm}[t]
\caption{Proposed User Association Algorithm} \label{Algorithm1}
\begin{algorithmic}[1]
\STATE \textbf{Input}: The power allocation satisfying \eqref{V1b} and the RRU activation satisfying \eqref{V1RRU}.
\FOR {$n \in \mathcal{N}$}
\STATE {Solve \eqref{ProblemAP3} to obtain the optimal user association for the $n$-th UE}.
\ENDFOR 
\STATE \textbf{Output}: The optimal user association vector $\boldsymbol{\mu}^*_n, \forall n\in \mathcal{N}$.
\end{algorithmic}
\end{algorithm} 

Given the variables of power allocation $\mathbf{p}$ and the RRU activation $\boldsymbol{\tau}$, the  user association problem is written as  
\begin{equation}\label{ProblemAP1}
    \begin{aligned}
        &&\underset{\substack{\boldsymbol{\Psi}}}{\mathrm{maximize}}& \quad \eta (\boldsymbol{\Psi},\boldsymbol{\tau},\mathbf{p})\\
        &&\mbox{s.t.}\quad& \quad \eqref{V1c}.
    \end{aligned}
\end{equation}
Notably, with fixed power allocation for all UEs and activation status of all RRUs, the total power consumption remains constant, regardless of the user association. Consequently, the optimization problem in \eqref{ProblemAP1} reduces to a total achievable rate maximization problem, expressed as

\begin{equation}\label{ProblemAP2}  
    \begin{aligned}
        && \underset{\substack{\boldsymbol{\Psi}}}{\mathrm{maximize}} & \hspace{0.2cm} R_{\textrm{tot}} \\
        &&\mbox{s.t.}\quad& \eqref{V1c}.
    \end{aligned}
\end{equation}
We first observe that the SINR value of the $n$-th UE, $\gamma_{n}$, in \eqref{SINR} depends solely on its own user association variable $\boldsymbol{\mu}_{n}$. Thus, the achievable rate of each UE is independent of the user association of other UEs. Consequently, the optimal user association for the $n$-th UE is determined by solving the following problem:
\begin{align}\label{ProblemAP3}  
    \begin{aligned}
        && \underset{\substack{\boldsymbol{\mu}_n}}{\mathrm{maximize}} & \quad B \log_2\left(1+\frac{\vert \boldsymbol{\mu}_{n}^T \boldsymbol{g}_n\vert ^2 P_{n}}{\underset{n'\neq n,\atop n'\in \mathcal{N}}{\sum}\vert \boldsymbol{\mu}_n^T \boldsymbol{g}_{n'}\vert ^2 P_{n'}+\sigma^2}\right) \\
        &&\mbox{s.t.}\quad&\quad \eqref{V1c}. 
    \end{aligned}
\end{align}
The proposed optimal user association algorithm is described in Algorithm~\ref{Algorithm1}.


\subsection{RRU Sensing Activation}\label{method:RRU_act}

\textcolor{black}{Given the variables of power allocation $\mathbf{p}$ and the user association $\boldsymbol{\Psi}$, the problem of RRU sensing activation is then written as}

\begin{align}\label{ProblemRRUV1}
    \begin{aligned}
        \underset{\substack{\boldsymbol{\tau}}}{\mathrm{maximize} } & \quad \eta(\boldsymbol{\Psi}, \mathbf{p}, \boldsymbol{\tau}) \\
        \mbox{s.t.}\quad &\quad  \eqref{V1RRU}, \eqref{P2:sense}.
    \end{aligned}
\end{align} 

When the allocated power values for all UEs and user association variables are fixed, the total achievable rate remains constant, irrespective of the RRU sensing activation. Therefore, the problem in \eqref{ProblemRRUV1} is reformulated as minimizing the number of activated RRUs as
\begin{align}\label{ProblemRRUV2}
    \begin{aligned}
        \underset{\substack{\boldsymbol{\tau}}}{\mathrm{minimize} } & \quad \sum_{k\in\mathcal{K}} \tau_{k}\\
        \mbox{s.t.}\quad &\quad \eqref{V1RRU}, \eqref{P2:sense} .
    \end{aligned}
\end{align}

To efficiently solve the binary RRU activation problem under sensing constraints, we design a decision-tree-based algorithm. Here, we introduce the CRLB sensing constraint value as follows
\begin{align}\label{eq:alg_RRU_CRLB}
    \Lambda(\boldsymbol{\tau}) = \mathbf{b}(\boldsymbol{\tau})  - \kappa \mathbf{A}(\boldsymbol{\tau})  \mathbf{p},
\end{align}
The process starts at the root node, where all RRUs are active, i.e., $\boldsymbol{\tau}^{\textrm{curr}} = [1, 1, \dots, 1]$. At each level of the decision tree, \textcolor{black}{we perform the following steps:}
\textit{(i)} Compute the current constraint value as in \eqref{eq:alg_RRU_CRLB}: $\Lambda^{\textrm{curr}} = \Lambda(\boldsymbol{\tau})$;
\textit{(ii)} For each active RRU $k$, temporarily deactivate it by setting $\tau_k = 0$ and compute the updated constraint: $\Lambda^{\textrm{temp}}_k = \Lambda(\boldsymbol{\tau}^{\textrm{temp}})$;
\textit{(iii)}  If the constraint remains satisfied (i.e., $\Lambda^{\textrm{temp}}_k \succ 0$), estimate the impact of this deactivation using the average degradation in the constraint. The RRU with the minimum impact is selected for permanent deactivation;
\textit{(iv)} Update the current activation vector by turning off the selected RRU, and move to the next level of the tree. 
This process repeats until no further RRU can be deactivated without violating the sensing constraint \eqref{P2:sense}, or only one RRU remains active. \textcolor{black}{The details of this decision tree-based RRU activation algorithm are provided in Algorithm~\ref{AlgorithmRRU}.}

\begin{algorithm}[t]
\caption{Decision Tree-Based RRU Activation Algorithm} \label{AlgorithmRRU}
\begin{algorithmic}[1]
\STATE \textbf{Input}: The power allocation $\mathbf{p}$ satisfying \eqref{V1b}.
\STATE \textbf{Initialize}: $\boldsymbol{\tau} \gets [1, 1, \dots, 1]$, $\mathcal{S} \gets \{1, 2, \dots, K\}$, $i \gets 0$
\WHILE{$|\mathcal{S}| > 1$}
    \STATE Compute the constraint $\Lambda^\textrm{curr}$ based on \eqref{eq:alg_RRU_CRLB}
    \FOR{$k \in \mathcal{S}$}
        \STATE Temporarily deactivate $k$: $\boldsymbol{\tau}^\textrm{temp} \gets \boldsymbol{\tau}, \tau_{k}^\textrm{temp} \gets 0$
        \STATE Compute $\Lambda^{\textrm{temp}}_{k} = \Lambda^\textrm{temp}(\boldsymbol{\tau}^{\textrm{temp}})$
        \IF{$\Lambda^\textrm{temp}_{k} \succ 0$}
            \STATE  $\Delta \Lambda_{k} \gets +\infty$
        \ELSE
            \STATE  $\Delta \Lambda_{k} \gets \textrm{mean}(\Lambda^\textrm{curr} - \Lambda^\textrm{temp}_{k}) $ 
        \ENDIF
    \ENDFOR
    \IF{$\Delta \Lambda_k = +\infty \textrm{ for all } k \in \mathcal{S}$}
        \STATE \textbf{break}
    \ENDIF
    \STATE $k^* \gets \arg\min_{k \in \mathcal{S}} \Delta \Lambda_k$
    \STATE $\boldsymbol{\tau}_{k^*} \gets 0$
    \STATE $\mathcal{S} \gets \mathcal{S} \setminus \{k^*\}$
\ENDWHILE
\STATE \textbf{Output}: The optimal RRU activation $\boldsymbol{\tau}^*\gets \boldsymbol{\tau}$.
\end{algorithmic}
\end{algorithm}

\begin{remark}
   Assuming all RRUs are initially active, the proposed algorithm iteratively deactivates one RRU at each step based on the sensing constraint. In the longest-run scenario, this process continues until only a single RRU remains active. The total computational complexity in this case is $\bigO(K^2)$, significantly lower than the exhaustive search method, which requires $\bigO(2^K)$ operations due to the binary activation states of each RRU, especially when the number of RRUs is large. This reduction in complexity makes the proposed approach more scalable and practical for larger networks.
\end{remark}

\subsection{Fixed-Point Power Allocation}\label{method:power}

Given the user association $\boldsymbol{\Psi}$ and the RRU activation $\boldsymbol{\tau}$, we can rewrite the achievable rate of the $n$-th UE as a function of power allocation $\mathbf{p}$, which is given by
\begin{equation}
R_n(\mathbf{p})=B\log(1+\gamma_{n}(\mathbf{p})).
\end{equation}
The problem in \eqref{Problemv2} is rewritten as
\begin{subequations}\label{ProblemP0}  
\begin{align} 
&& \underset{\substack{ \mathbf{p}}}{\mathrm{maximize}} & \hspace{0.2cm} \eta(\mathbf{p}) \\
&&\mbox{s.t.}\quad&\hspace{0.2cm} 0 \leq P_{n} \leq P^{\textrm{max}}_{n} \hspace{0.3cm} ,\forall n\in \mathcal{N} ,\\
&&& \hspace{0.2cm} \eqref{P2:sense},\nonumber
\end{align}
\end{subequations} 
where $\eta(\mathbf{p})$ is calculated as \eqref{EEfirst} with the fixed user association $\boldsymbol{\Psi}$ and the fixed RRU activation $\boldsymbol{\tau}$. Notably, the problem in \eqref{ProblemP0} is a non-linear fractional program and hard to obtain the optimal solution. Our approach is to provide a tractable form of the problem by considering the log transformation \cite{Singh2015} as 
\begin{equation}\label{LogRelax}
\log_2 (1+z) \geq  \alpha \log_2 (z) + \beta,
\end{equation}
where the log transformation factors $\alpha$ and $\beta$ are given as 
\begin{equation}
\begin{aligned}\label{LogApproximation}
&\alpha=\frac{z}{1+z},\\
&\beta=\log_2(1+z)-\frac{z}{1+z}\log_2(z).
\end{aligned} 
\end{equation}
Here, $z$ can be any non-negative value. We apply the transformation in \eqref{LogRelax} to the achievable rate $R_{n}$ in \eqref{RateCalculation} as
\begin{equation}
R_{n}(\mathbf{p}) \geq \tilde{R}_{n}(\mathbf{p}) = B\left(\alpha_{n}(\mathbf{p})\log_2 (\gamma_{n}(\mathbf{p}))+\beta_{n}(\mathbf{p})\right),
\end{equation}
where the following definitions hold
\begin{equation}
\begin{aligned}\label{LogApproximation1}
&\alpha_{n}(\mathbf{p})=\frac{\gamma_{n}(\mathbf{p})}{1+\gamma_{n}(\mathbf{p})},\\
&\beta_{n}(\mathbf{p})=\log_2(1+\gamma_{n}(\mathbf{p}))-\frac{\gamma_{n}(\mathbf{p})}{1+\gamma_{n}(\mathbf{p})}\log_2(\gamma_{n}(\mathbf{p})).
\end{aligned} 
\end{equation}
We define $\boldsymbol{\alpha}(\mathbf{p})=\{\alpha_{n}(\mathbf{p})\}\in \mathbb{R}^{N}$ and $\boldsymbol{\beta}(\mathbf{p})=\{\beta_{n}(\mathbf{p})\}\in \mathbb{R}^{N}$. The transformation of the EE after applying the log transformation is given by
\begin{align} \label{LowerboundInitial}
\begin{aligned}
\tilde{\eta}(\mathbf{p}) &= \tilde{\eta
}(\mathbf{p},\boldsymbol{\alpha}(\mathbf{p}),\boldsymbol{\beta}(\mathbf{p})) \\
&= \frac{\tilde{R}_{n}(\mathbf{p})}{\sum_{n\in \mathcal{N}}(P_{\textrm{cir}}+P_{n}) + P^{\textrm{con,RRU}}},
\end{aligned}
\end{align}
Then the problem in \eqref{ProblemP0} is approximated as follows 
\begin{subequations}\label{ProblemP01}  
\begin{align} 
&& \underset{\substack{ \mathbf{p}}}{\mathrm{maximize}} & \hspace{0.2cm} \tilde{\eta}(\mathbf{p}) \\
&&\mbox{s.t.}\quad&\hspace{0.2cm} 0 \leq P_{n} \leq P^{\textrm{max}}_{n} \hspace{0.3cm} ,\forall n\in \mathcal{N} \\
&&& \hspace{0.2cm} \eqref{P2:sense}\nonumber.
\end{align}
\end{subequations} 
To solve this problem, we propose an iterative algorithm for approximation solution. Particularly, there are two main steps including: $(i)$ optimizing $\tilde{\eta}({\mathbf{p}})$ while the log transformation factors $\boldsymbol{\alpha}(\mathbf{p})$ and $\boldsymbol{\beta}(\mathbf{p})$  are fixed, then $(ii)$ updating the aforementioned log transformation factors based on the new power variable $\mathbf{p}$. 
Introduce the new power notation at this algorithm $ \mathbf{p}_{\textrm{Log}}^{(0)} = \mathbf{p}^{(t-1)}$. At the $i$-th iteration of the procedure, we solve the following problem 
\begin{align}\label{ProblemP_Algorithm}  
&& \underset{\substack{ \mathbf{p}_{\textrm{Log}}^{(i)}}}{\mathrm{maximize}} & \quad \frac{\displaystyle\sum_{n\in \mathcal{N}}B \left(\alpha_{n}(\mathbf{p}_{\textrm{Log}}^{(i-1)})\log (\gamma_{n}(\mathbf{p}_{\textrm{Log}}^{(i)}))+\beta_{n}(\mathbf{p}_{\textrm{Log}}^{(i-1)})\right)}{\sum_{n\in \mathcal{N}}(P_{\textrm{cir}}+P_{\textrm{Log},n}^{(i)}) + P^{\textrm{con,RRU}}} \nonumber\\
&&\mbox{s.t.}\quad &\quad 0 \leq P_{\textrm{Log},n}^{(i)} \leq P^{\textrm{max}}_{n} \quad ,\forall n\in \mathcal{N}, \\
&&&\quad  \mathbf{b} - \kappa \mathbf{A} \mathbf{p}_{\textrm{Log}}^{(i)} \preceq 0, \nonumber
\end{align}
\hspace{-0.2cm}
where $\mathbf{p}_{\textrm{Log}}^{(i-1)^*}$ is the power allocation vector obtained from the $(i-1)$-th iteration. 
The logarithmic approximation method is terminated when the variation between two consecutive iterations is smaller than a sufficiently small tolerance $\tilde{\eta}(\mathbf{p}_{\textrm{Log}}^{(i)^*})~-~\tilde{\eta}(\mathbf{p}_{\textrm{Log}}^{(i-1)^*})<\epsilon_1 $. The detailed procedure solving problem \eqref{ProblemP_Algorithm} is provided as in Algorithm \ref{LogApproximationAlgorithm}, and its convergence is proven in the following Lemma. 
\begin{algorithm}[t]
\caption{Proposed solution to problem \eqref{ProblemP_Algorithm}} \label{LogApproximationAlgorithm} 
\begin{algorithmic}[1] 
\STATE \textbf{Input}: The user association $\boldsymbol{\Psi}$, the RRU activation $\boldsymbol{\tau}$, and the  power allocation vector $\mathbf{p}^{(t)}$.
\STATE Set $i=0$ $\mathbf{p}_{\textrm{Log}}^{(0)} = \mathbf{p}^{(t)}$
\STATE Calculate $\boldsymbol{\alpha}(\mathbf{p}_{\textrm{Log}}^{(0)})$ and $\boldsymbol{\beta}(\mathbf{p}_{\textrm{Log}}^{(0)})$. 
\REPEAT 
\STATE {$i\gets i+1$.} 
\STATE {Solve the problem \eqref{ProblemP_Algorithm} with fixed $\boldsymbol{\alpha}(\mathbf{p}_{\textrm{Log}}^{(i-1)^*})$ and $\boldsymbol{\beta}(\mathbf{p}_{\textrm{Log}}^{(i-1)^*})$ to obtain the solution $\mathbf{p}_{\textrm{Log}}^{(i)^*}$.} 
\STATE {Update $\boldsymbol{\alpha}(\mathbf{p}_{\textrm{Log}}^{(i)^*})$ and $\boldsymbol{\beta}(\mathbf{p}_{\textrm{Log}}^{(i)^*})$.}
\UNTIL {$\tilde{\eta}(\mathbf{p}_{\textrm{Log}}^{(i)^*})-\tilde{\eta}(\mathbf{p}_{\textrm{Log}}^{(i-1)^*})<\epsilon_1 $} 
\STATE \textbf{Output}: The power allocation vector. 
\end{algorithmic}
\end{algorithm} 

\begin{lemma}\label{LogLemma}
Iteratively solving the power allocation problem in \eqref{ProblemP_Algorithm} and updating the log transformation factors as Algorithm \ref{LogApproximationAlgorithm} will monotonically increase the objective function of \eqref{ProblemP0}, which always converges.
\end{lemma}
\begin{proof} 
The proof is provided in Appendix \ref{ProofLemmaLogApproximation}.
\end{proof}

To solve the problem in \eqref{ProblemP_Algorithm}, we exploit the Dinkelbach procedure \cite{Dinkelbach1967} as a canonical method to transform the problem in \eqref{ProblemP_Algorithm} into a series of parametric subtractive-form. 
We initialize the new notation for power variable in this Dinkelbach procedure,  
$\mathbf{p}_\textrm{Din}^{(0)}=\mathbf{p}_{\textrm{Log}}^{(i-1)^*}$. At the $r$-th iteration, we solve the following problem 
\begin{equation}\label{ProblemP2}  
\begin{aligned}
&& \underset{\substack{ \mathbf{p}_{\textrm{Din}} }}{\mathrm{maximize}} & \quad \sum_{n\in \mathcal{N}}\tilde{R}_n(\mathbf{p}_{\textrm{Din}})-\tilde{\eta}_{\textrm{Din}}^{(r-1)^*}\sum_{n\in \mathcal{N}}P_{\textrm{Din,n}}^{\textrm{con}} \\
&&\mbox{s.t.}\quad&\quad 0 \leq P_{\textrm{Din},n} \leq P^{\textrm{max}}_{n} \quad ,\forall n\in \mathcal{N},\\ 
&&&\quad  \mathbf{b} - \kappa \mathbf{A} \mathbf{p}_{\textrm{Din}} \preceq 0,
\end{aligned}
\end{equation} 
where $P_{\textrm{Din,n}}^{\textrm{con}} = P_{\textrm{Din,n}} + P_{\textrm{cir}}, \forall n\in\mathcal{N}$. Following the Dinkelbach procedure, we define $\tilde{\eta}_{\textrm{Din}}^{(r-1)^*}=\tilde{\eta}(\mathbf{p}_{\textrm{Din}}^{(r-1)^*})$ as in \eqref{LowerboundInitial}, and get $\mathbf{p}_{\textrm{Din}}^{(r)^*}$ as the solution at the $r$-th iteration is  $\mathbf{p}_{\textrm{Din}}^{(r)^*}$.
\begin{figure*}[t]
\begin{align}\label{ProblemP3}  
&& \underset{\substack{  \bar{\mathbf{p}}_\textrm{Din} }}{\mathrm{maximize}} & \quad \sum_{n\in \mathcal{N}} B \left[\alpha_{n}\log_2\left(\frac{\vert \boldsymbol{\mu}_{n}^T \boldsymbol{g}_n\vert ^2 \exp(\bar{P}_{\textrm{Din},n})}{\displaystyle\sum_{n'\in \mathcal{N}\setminus \{n\}}\vert \boldsymbol{\mu}_n^T \boldsymbol{g}_{n'}\vert ^2\exp(\bar{P}_{\textrm{Din},n'})+\sigma^2}\right)+\beta_{n}\right] -\tilde{\eta}_{\textrm{Din}}^{(r-1)^*} \left( P^{\textrm{con,RRU}}+\sum_{n\in \mathcal{N}} (P_{\textrm{cir}}+\exp (\bar{P}_{\textrm{Din},n}))\right)  \nonumber\\
&&\mbox{s.t.}\quad&\quad \exp(\bar{P}_{\textrm{Din},n}) \leq P^{\textrm{max}}_{n} \quad ,\forall n\in \mathcal{N} , \\
&&& \quad  \mathbf{b} - \kappa \mathbf{A}\exp(\bar{\mathbf{p}}_\textrm{Din})\leq 0. \nonumber
\end{align} 
\hrule
\vspace*{-0.6cm}
\end{figure*}
According to \cite{Dinkelbach1967}, the Dinkelbach method ensures that the iterative solution of problem \eqref{ProblemP2} will converge to the solution \eqref{ProblemP_Algorithm}. 
We observe that problem \eqref{ProblemP2} has a hidden-convex form. Thus, we apply the transformation $P_{\textrm{Din},n}=\exp(\bar{P}_{\textrm{Din},n}), \forall n\in \mathcal{N}$, and $\bar{\mathbf{p}}_\textrm{Din}=[\bar{P}_{\textrm{Din},1},...,\bar{P}_{\textrm{Din},N}]$. The problem in \eqref{ProblemP2} is rewritten with the new variable $\bar{P}_{\textrm{Din},n}$ as presented in \eqref{ProblemP3} at the top of next page. 
Here, we verify the convexity of the problem in \eqref{ProblemP3}. It is observed that the second term in the objective function, $\tilde{\eta}_{\textrm{Din}}^{(r-1)^*} \sum_{n\in \mathcal{N}}(P_{\textrm{cir}}+\exp (\bar{P}_{\textrm{Din},n}))$, is always convex. The first term in the objective function is rewritten as follows
\begin{equation}\label{SumRate}
\begin{aligned}
&\sum_{n\in \mathcal{N}} B \Bigg[\alpha_{n}\Bigg(\log_2\left(\vert \boldsymbol{\mu_{n}}^T\boldsymbol{g}_n\vert ^2\exp(\bar{P}_{\textrm{Din},n})\right)-\\
&\log_2\bigg(\sum_{^{n'\neq n}_{n'\in \mathcal{N}}}\vert \boldsymbol{\mu}_n^T\boldsymbol{g}_{n'}\vert ^2\exp(\bar{P}_{\textrm{Din},n'})+\sigma^2\bigg)\Bigg) +\beta_{n}\Bigg].
\end{aligned}
\end{equation} 
First, $\log_2\left(\vert\boldsymbol{\mu_{n}}^T\boldsymbol{g}_{n}\vert^2\exp(\bar{P}_{\textrm{Din},n})\right)$ is an affine function. Meanwhile, $\log_2(\sum_{n'\in \mathcal{N}\setminus \{n\} }\vert\boldsymbol{\mu}_n^T\boldsymbol{g}_{n'}\vert^2\exp(\bar{P}_{\textrm{Din},n'})+\sigma^2)$ is the log-sum-exp function, and thus it is convex \cite{Boyd2004}. Therefore, $\log_2\left(\vert\boldsymbol{\mu_{n}}^T\boldsymbol{g}_{n}\vert^2\exp(\bar{P}_{\textrm{Din},n})\right)-\log_2(\sum_{n'\in \mathcal{N}\setminus \{n\}} \vert\boldsymbol{\mu}_n^T\boldsymbol{g}_{n'}\vert^2\exp(\bar{P}_{\textrm{Din},n'})+\sigma^2)$ is concave on $\bar{P}_{\textrm{Din},n}$. Consequently, the sum in \eqref{SumRate} is a concave function on new variables $\bar{P}_{\textrm{Din},n}$, and thus the objective function of \eqref{ProblemP3} is concave on $\bar{P}_{\textrm{Din},n}$. Besides, all the constraints of the problem in \eqref{ProblemP3} are affine. Therefore, we can conclude that the problem \eqref{ProblemP3} is a convex optimization problem. 
\begin{figure*}[t]
\begin{align}\label{LagrangianFunction}
\mathcal{L}(\bar{\mathbf{p}}_\textrm{Lag},\boldsymbol{\lambda},\boldsymbol{\nu})&=\sum_{n\in \mathcal{N}} B \Bigg[\alpha_{n,l}\times \log_2\left(\frac{\vert\boldsymbol{\mu_{n}}^T\boldsymbol{g}_n\vert^2\exp (\bar{P}_{\textrm{Lag},n})}{\sum_{n'\in \mathcal{N}\setminus \{n\}} \vert\boldsymbol{\mu}_n^T\boldsymbol{g}_{n'}\vert^2\exp (\bar{P}_{\textrm{Lag},n'})+\sigma^2}\right) +\beta_{n}\Bigg] 
- \tilde{\eta}_{\textrm{Din}}^{(r-1)^*} \left( P^{\textrm{con,RRU}} + \sum_{n\in \mathcal{N}}(P_{\textrm{cir}}+\exp(\bar{P}_{\textrm{Lag},n})) \right) \nonumber\\
&  -\sum_{n\in \mathcal{N}}\lambda_n(\exp(\bar{P}_{\textrm{Lag},n})-P_n^{\textrm{max}})  - \sum_{n \in \mathcal{N}}\nu_{n} \biggl(\kappa \mathbf{a}_{n} \exp(\bar{\mathbf{p}}_\textrm{Lag}) - b_{n} \biggl). 
\end{align}
\hrule
\vspace*{-0.25cm}
\end{figure*}
At this step, we suggest applying the Lagrangian method to obtain the solution of the problem in \eqref{ProblemP3}. We use the obtained power from the Dinkelbach procedure as the initial input of the Lagrangian procedure $\mathbf{p}_{\textrm{Lag}}^{(0)}=\mathbf{p}_{\textrm{Din}}^{(r-1)^*}$. The Lagrangian function is given by \eqref{LagrangianFunction} at the top of this page,
where $\lambda_n$ and $\nu_n$ are the Lagrangian dual variables corresponding to the maximum power constraint and sensing CRLB requirement of the $n$-th UE, respectively. \textcolor{black}{We define $\boldsymbol{\lambda}=[\lambda_1,...,\lambda_N]$, $\boldsymbol{\nu}=[\nu_1,...,\nu_N]$, $\mathbf{b} = [b_1, \dots, b_N]^T$, $\mathbf{A} = [\mathbf{a}_1, \dots, \mathbf{a}_N]^T$. }
Based on the Lagrangian function \eqref{LagrangianFunction}, the dual optimization problem is given as
\begin{equation}\label{DualProblem}  
\begin{aligned}
&& \underset{\substack{\boldsymbol{\lambda}, \boldsymbol{\nu} }}{\mathrm{minimize}} & \quad g(\boldsymbol{\lambda}, \boldsymbol{\nu})=\underset{\bar{\mathbf{p}}_\textrm{Lag}}{\mathrm{maximize}} \hspace{0.1cm}\mathcal{L}(\bar{\mathbf{p}}_\textrm{Lag},\boldsymbol{\lambda},\boldsymbol{\nu}).\\
&&\mbox{s.t.}\quad&\quad \lambda_n \geq 0 \quad ,\forall n\in \mathcal{N}, \\
&&&\quad \nu_n \geq 0 \quad ,\forall n\in \mathcal{N}.
\end{aligned} 
\end{equation} 
Since problem \eqref{ProblemP3} is convex, with all constraints forming sub-level sets of convex functions and being strictly feasible, Slater's condition holds. Consequently, there is no duality gap \cite{Boyd2004}, allowing the problem to be efficiently solved via its dual formulation in \eqref{DualProblem}. 
We suggest an iterative procedure to solve the dual problem in \eqref{DualProblem}. Specifically, we optimize $\mathcal{L}(\bar{\mathbf{p}}_\textrm{Lag},\boldsymbol{\lambda}, \boldsymbol{\nu})$ with respect to $\bar{\mathbf{p}}_\textrm{Lag}$. Then we update the Lagrangian multipliers using the sub-gradient method to penalize the violation of limited power constraints as follows
\begin{align}
    \begin{aligned}\label{LagranMultiplierUpdate}
\lambda_n^{(s+1)}&=\mathrm{max} (0,\lambda_n^{(s)}-\varepsilon (P^{\textrm{max}}-P_{\textrm{Lag},n})), \\
\nu_n^{(s+1)}&=\mathrm{max} \left(0,\nu_n^{(s)}-\varepsilon \left(\kappa \mathbf{a}_{n} \mathbf{p}_\textrm{Lag} - b_n\right)\right),
\end{aligned}
\end{align}
where $\varepsilon$ is a sufficiently small step size. 
\textcolor{black}{To find the power that maximizes the Lagrangian function in \eqref{LagrangianFunction}, we apply a fixed-point procedure.} We use the obtained power from the Lagrangian procedure as the initial input power allocation of this procedure $\mathbf{p}_{\textrm{SIF}}^{(0)}=\mathbf{p}_{\textrm{Lag}}^{(s-1)}$.
To this end, the corresponding derivative of $\mathcal{L}(\bar{\mathbf{p}}_\textrm{SIF},\boldsymbol{\lambda},\boldsymbol{\nu})$ with respect to $\bar{P}_{\textrm{SIF},n}$ is given as
\begin{equation}\label{KKTconditions} 
\begin{aligned}
&\frac{\partial \mathcal{L}}{\partial \bar{P}_{\textrm{SIF},n}}=B \alpha_{n}\log_{2}(\textrm{e})-\sum_{j\in \mathcal{N}\setminus \{n \}}B\alpha_{j} \\
&\times \frac{\vert\boldsymbol{\mu}^T_j \boldsymbol{g}_n\vert^2 \exp (\bar{P}_{\textrm{SIF},n})}{\ln (2)\bigg(\sum_{n'\in \mathcal{N}\setminus \{j \}} \vert\boldsymbol{\mu}_j^T\boldsymbol{g}_{n'}\vert^2\exp(\bar{P}_{\textrm{SIF},n'})+\sigma^2\bigg)}\\
&-(\lambda_n+\tilde{\eta}_{\textrm{Din}}^{(r-1)^*} + \nu_n \kappa A_{n,n})\exp (\bar{P}_{\textrm{SIF},n}),
\end{aligned}
\end{equation} 
where $\mathbf{a}_{n} = [A_{1,n}, \cdots, A_{N,n}]$.
\begin{figure*}[t]
\begin{equation}\label{PowerUpdate}
P_{\textrm{SIF},n}=\frac{B\alpha_{n }\log_{2}(\textrm{e})}{\sum_{j\in \mathcal{N}\setminus \{n\}}B\alpha_{j}\frac{\vert\boldsymbol{\mu}_j^T \boldsymbol{g}_n\vert^2}{\ln (2) \left(\sum_{n'\in \mathcal{N}\setminus \{j\}}\vert\boldsymbol{\mu}_{j}^T \boldsymbol{g}_{n'}\vert^2 P_{\textrm{SIF},n'}+\sigma^2\right)}+\lambda_n+\tilde{\eta}_{\textrm{Din}}^{(r-1)^*} + \nu_n \kappa A_{n,n}}=I_n(\mathbf{p}_\textrm{SIF}).
\end{equation}
\hrule
\vspace*{-0.25cm}
\end{figure*}
\noindent
From \eqref{KKTconditions}, we back-transform $\exp (\bar{P}_{\textrm{SIF},n}) = P_{\textrm{SIF},n}$, 
a set of fixed-point solutions to the power allocation is formed as in \eqref{PowerUpdate}. We define the right-hand side of the equation \eqref{PowerUpdate} as $I_n(\mathbf{p}_\textrm{SIF})$, and we have 
\begin{equation}\label{FixedPointEquation}
P_{\textrm{SIF},n}=I_n(\mathbf{p}_\textrm{SIF}).
\end{equation}
We define $\boldsymbol{I}(\mathbf{p}_\textrm{SIF})=[I_1(\mathbf{p}_\textrm{SIF}), ..., I_N(\mathbf{p}_\textrm{SIF})]$. The power vector will be updated as the following iterative algorithm
\begin{equation}\label{FixedPointAlgorithm}
\mathbf{p}_\textrm{SIF}^{(m+1)}=\boldsymbol{I}(\mathbf{p}_\textrm{SIF}^{(m)}),
\end{equation}
where $m$ is the iteration index.
The power $\boldsymbol{I}(\mathbf{p}_\textrm{SIF}^{(m)})$ is the obtained power from previous iteration. The convergence of \eqref{FixedPointAlgorithm} is proved by showing that $\boldsymbol{I}(\mathbf{p}_\textrm{SIF})$ is a standard interference function \cite{Yates1995}. We can provide the following result on the fixed-point equation in \eqref{FixedPointAlgorithm}.

\begin{lemma}\label{SIFlemma}
Assume that power allocation $\normalfont{\mathbf{p}}_\textrm{SIF}$ always satisfies $0\preceq \normalfont{\mathbf{p}}_\textrm{SIF} \preceq \normalfont{\mathbf{p}}^{\textrm{max}}$, then the function $\mathbf{p}_\textrm{SIF}=\boldsymbol{I}(\mathbf{p}_\textrm{SIF})$ with its elements as given in \eqref{PowerUpdate} is a standard interference function (SIF). Thus, the iterative update based on \eqref{FixedPointAlgorithm} converges to a fixed point that optimize the problem \eqref{ProblemP2}.    
\end{lemma}
\begin{proof}
The proof is given in Appendix \ref{ProofLemma3}. 
\end{proof}

After obtaining the power from the fixed-point algorithm in \eqref{FixedPointAlgorithm}, we update the Lagrangian multipliers until convergence. Then, we achieve the power that maximizes the Lagrangian function in \eqref{LagrangianFunction}. The achieved power is then used to update in the Dinkelbach procedure. The Dinkelbach procedure converges when $\sum_{n\in \mathcal{N}}{R}_n(\mathbf{p}_{\textrm{Din}}^{(r)}) -\tilde{\eta}_{\textrm{Din}}^{(r-1)^*} \sum_{n\in \mathcal{N}} P_{\textrm{Din},n}^{\textrm{con},(r)}$ approaches zero. The final power vector from the Dinkelbach procedure is the optimal power that maximizes $\tilde{\eta}_{\textrm{Din}}$. This power vector is then used to update the power allocation for the logarithmic approximation algorithm  of Algorithm \ref{LogApproximationAlgorithm}. Finally, we summarize the process to solve the problem in \eqref{ProblemP_Algorithm} as in Algorithm \ref{Algorithm3}.

\begin{algorithm}[t]
\caption{Proposed fixed-point power allocation algorithm to solve \eqref{ProblemP_Algorithm}.} \label{Algorithm3}
\begin{algorithmic}[1]
\STATE \textbf{Input}: $\boldsymbol{\Psi}$, and $\mathbf{p}_{\textrm{Log}}^{(i)}$.
\STATE Set $r=0$, $\mathbf{p}_{\textrm{Din}}^{(0)}=\mathbf{p}_{\textrm{Log}}^{(i)}$, and compute $\tilde{\eta}_{\textrm{Din}}^{(r)^*}=\frac{\sum_{n\in \mathcal{N}} \tilde{R}_{n}(\mathbf{p}_{\textrm{Din}}^{(r)})}{P^{\textrm{con,RRU}}+\sum_{n\in \mathcal{N}} P^{\textrm{con},(r)}_{\textrm{Din},n}}$.
\REPEAT 
\STATE Set $s=0$, $\mathbf{p}_{\textrm{Lag}}^{(0)}=\mathbf{p}_{\textrm{Din}}^{(r)}$, and initiate $\boldsymbol{\lambda}^{(0)}, \boldsymbol{\nu}^{(0)}$.
\REPEAT
\STATE Set $m=0$ and $\mathbf{p}_{\textrm{SIF}}^{(0)}=\mathbf{p}_{\textrm{Lag}}^{(s)}$.
\REPEAT 
\STATE {Update $\mathbf{p}_{\textrm{SIF}}^{(m+1)}$ using \eqref{FixedPointAlgorithm} and update $m=m+1$.}
\UNTIL {$\mathbf{p}_{\text{SIF}}$ converge}. \textbf{Return} $\mathbf{p}_{\textrm{Lag}}^{(s+1)}$.
\STATE {Update $\boldsymbol{\lambda}^{(s+1)}$ and $\boldsymbol{\nu}^{(s+1)}$ using \eqref{LagranMultiplierUpdate} and update $s=s+1$.}
\UNTIL {$\mathbf{\lambda}, \mathbf{\nu}$ converge}. \textbf{Return} $\mathbf{p}_{\textrm{Din}}^{(r+1)^*}$
\STATE {Compute $\Delta=\sum_{n\in \mathcal{N}}{R}_n(\mathbf{p}_{\textrm{Din}}^{(r+1)}) -\tilde{\eta}^{(r)} \sum_{n\in \mathcal{N}} P_{\textrm{Din},n}^{\textrm{con},(r+1)}$.}
\STATE {Update:} $\tilde{\eta}_{\textrm{Din}}^{(r+1)^*}=\frac{\sum_{n\in \mathcal{N}} R_{n}(\mathbf{p}_{\textrm{Din}}^{(r+1)^*})}{P^{\textrm{con,RRU}}+\sum_{n\in \mathcal{N}} P^{\textrm{con},(r+1)^*}_{\textrm{Din},n}}$.
\STATE $r=r+1$.
\UNTIL {$ \Delta < \epsilon_2$}
\STATE \textbf{Output}: The power allocation vector $\mathbf{p}_{\textrm{Log}}^{(i+1)}$. 
\end{algorithmic} 
\end{algorithm}

\subsection{Alternative Algorithm} 

To solve the optimization problem in \eqref{Problemv2}, we propose an Alternative Algorithm (AA algorithm) that iteratively optimizes user association, RRU activation, and power allocation until convergence in EE value.
The algorithm initializes the power allocation vector and RRU activation vector as $\mathbf{p}^{(0)}$ and $\boldsymbol{\tau}^{(0)}$, respectively. At each iteration $i$, the three sub-problems are solved sequentially using the latest updates from the previous step. First, user association $\boldsymbol{\Psi}^{(i)}$ is updated by solving \eqref{ProblemAP1} using Algorithm~\ref{Algorithm1}. Next, RRU activation $\boldsymbol{\tau}^{(i)}$ is optimized via the decision tree-based approach in Algorithm~\ref{AlgorithmRRU} for \eqref{ProblemRRUV2}. Finally, power allocation $\mathbf{p}^{(i)}$ is determined using the fixed-point algorithm in Algorithm~\ref{LogApproximationAlgorithm} for \eqref{ProblemP0}.
The algorithm terminates when the EE improvement falls below a predefined threshold, i.e., $\eta\big(\boldsymbol{\Psi}^{(t)},\mathbf{p}^{(t)}, \boldsymbol{\tau}^{(t)}\big)-\eta\big(\boldsymbol{\Psi}^{(t-1)},\mathbf{p}^{(t-1)}, \boldsymbol{\tau}^{(t-1)}\big)<\epsilon_{3}$. The detailed procedure is outlined in Algorithm~\ref{AAlgorithm}, and its convergence is established in the following lemma.

\begin{algorithm}[t]
\caption{The Alternative Algorithm} \label{AAlgorithm}
\begin{algorithmic}[1] 
\STATE \textbf{Input}: Initiate power allocation $\mathbf{p}^{(0)}=\mathbf{p}^{\textrm{max}}$, RRU activation $\mathbf{\tau}^{(0)}$, and the algorithm stopping tolerance $\epsilon_3 > 0$. 
\STATE Set the iteration index $t=0$.
\REPEAT 
\STATE Set $t=t+1$.
\STATE Given $\mathbf{p}^{(t-1)}$ and $\boldsymbol{\tau}^{(t-1)}$, solve the problem in \eqref{ProblemAP1} to obtain user association $\boldsymbol{\Psi}^{(t)}$ 
\STATE Given $\boldsymbol{\Psi}^{(t)}$ and $\mathbf{p}^{(t-1)}$, solve the problem in \eqref{ProblemRRUV2} to obtain better RRU activation $\boldsymbol{\tau}^{(t)}$
\STATE Given $\boldsymbol{\Psi}^{(t)}$ and $\boldsymbol{\tau}^{(t)}$, solve the problem in \eqref{ProblemP0} to obtain better power allocation $\mathbf{p}^{(t)}$
\UNTIL $\eta(\boldsymbol{\Psi}^{(t)},\boldsymbol{\tau}^{(t)},\mathbf{p}^{(t)})-\eta(\boldsymbol{\Psi}^{(t-1)},\boldsymbol{\tau}^{(t-1)},\mathbf{p}^{(t-1)})<\epsilon_{3}$ 
\STATE \textbf{Output}: The optimal user association $\boldsymbol{\Psi}^{*}$, the RRU activation $\boldsymbol{\tau}^{*}$, and the transmit power $\mathbf{p}^{*}$.
\end{algorithmic}
\end{algorithm}

\begin{lemma}\label{AlternativeLemma}
Assume that the initial power allocation $\normalfont{\mathbf{p}}^{(0)}$ and RRU activation $\boldsymbol{\tau}^{(0)}$ of Algorithm \ref{AAlgorithm} satisfy \eqref{V1c} and \eqref{V1RRU}, respectively. Using Algorithm \ref{AAlgorithm}, the EE value, $\eta(\boldsymbol{\Psi}^{(t)},\normalfont{\mathbf{p}}^{(t)},\boldsymbol{\tau}^{(t)})$, is non-decreasing at the $t$-th iteration, and this value always converges.
\end{lemma}

\begin{proof}
The proof is given in Appendix \ref{ProofLemma4}.
\end{proof}

\begin{remark}
In the multi-target setting, the FIM is a block-structured extension of the single-target case and remains affine in the uplink power allocation.
Using Schur complements, the CRLB constraints can be written as linear matrix inequalities (LMIs), guaranteeing convex feasibility of the power allocation sub-problem in (20). Hence, alternative algorithm Algorithm~\ref{AAlgorithm} remains applicable to multi-target scenarios without modification.
\end{remark}

\subsection{Complexity Analysis}

The complexity of the AA algorithm in solving problem~\eqref{Problemv2} mainly arises from the iterative nature of its three sub-problems and the outer iteration loop. \textcolor{black}{In particular, the complexity of the user association method in Algorithm~\ref{Algorithm1} is on the order of $\bigO(K N^{2})$, the proposed RRU activation method in Algorithm~\ref{AlgorithmRRU} has a complexity of $\bigO(K^2)$, as discussed in Section~\ref{method:RRU_act}. }
Finally, the fixed-point power allocation in Algorithm~\ref{LogApproximationAlgorithm} has a complexity of $\bigO(D^{\textrm{SIF}} D^{\textrm{Lag}} D^{\textrm{Din}} D^{\textrm{Log}} N)$, where $D^{\textrm{SIF}}, D^{\textrm{Lag}}, D^{\textrm{Din}}$, and $D^{\textrm{Log}}$ denote the number of iterations required for convergence of the fixed-point update, Lagrangian multipliers, Dinkelbach’s method, and the logarithmic approximation, respectively.
Assuming that Algorithm~\ref{AAlgorithm} requires $D^{\textrm{AA}}$ iterations to converge, the overall complexity of the model-based algorithm is on the order of
\begin{align}\label{eq:complex_model}
    \bigO\left( D^{\textrm{AA}} \left(KN^{2} + K^{2} + D^{\textrm{SIF}} D^{\textrm{Lag}} D^{\textrm{Din}} D^{\textrm{Log}} N \right) \right).
\end{align}
However, the actual runtime depends heavily on network scale and convergence thresholds, and may therefore vary across different system settings.

\section{Graph Neural Network Approach}\label{sec:ISAC-GNN}

This section provides a DL framework for the resource allocation problem in the considered ISAC system. First, the C-RAN ISAC system is represented as a heterogeneous graph. Next, we propose a GNN architecture, which learns the joint resource management policy for the aforementioned ISAC system. The detailed design of the GNN is then discussed.

\subsection{Graph Modeling the ISAC System} \label{sec:graph_model}
We model the ISAC C-RAN system as an undirected bipartite graph, denoted as $\mathcal{G} = \{\mathcal{V}^{\textrm{u}}, \mathcal{V}^{\textrm{r}}, \mathcal{E}\}$, where $\mathcal{V}^{\textrm{u}}$, $\mathcal{V}^{\textrm{r}}$, and $\mathcal{E}$ represent the sets of UE nodes, RRU nodes, and edges between them, respectively. This structure, referred to as the Integrated Sensing and Communication Graph (ISCG), captures the interaction between UEs and RRUs, where each UE node is connected to all RRUs. The nodes and edges are characterized by the following features:
\begin{itemize}
    \item UE nodes $\mathcal{V}^\textrm{u}$: We denote the features of UE node $v^{\textrm{u}}_{n}$ as $u_{n}$. This node includes the maximum power threshold $P^{\textrm{max}}_{n}$, the circuit power consumption $P^{\textrm{cir}}$, the localization accuracy threshold $\kappa$, and the corresponding 2-D coordinate $(\hat{x}_{n}, \hat{y}_{n})$. 
    \item RRU nodes $\mathcal{V}^\textrm{r}$: We denote the feature of $v^{\textrm{r}}_{k}$ as $r_{k}$. \textcolor{black}{The features of RRU node include the corresponding 2-D coordinate $(\tilde{x}_{k}, \tilde{y}_{k})$, and the sensing power consumption $P_{\textrm{sens}}$. 
    \item UE-RRU edges $\mathcal{E}$: We denote the attribute of the edge linking $v^{\textrm{u}}_{n}$ and $v^{\textrm{r}}_{k}$  as $e_{kn}$, whose attributes are the channel gain $g_{kn}$, and the RCS $\rho_{kn}$ of the UE-target-RRU path.  }   
\end{itemize}
Fig.~\ref{fig:graph_present} illustrates the ISCG representing the considered system with $N = 3$ UEs and $K = 2$ RRUs. This graph representation effectively captures the permutation equivariance property of the resource allocation problem in \eqref{Problemv1}, ensuring that reordering UE or RRU nodes results in a corresponding adjustment in the solution. Conventional ML and DL models fail to exploit this graph-structured interaction, lacking permutation equivariance. To address this challenge, we design a GNN that leverages the inherent graph structure of the ISCG for efficient resource allocation.

\subsection{Graph Neural Networks for Integrated Sensing and Communication Networks}
\begin{figure*}[t]
    \begin{minipage}{0.3\textwidth}
    \includegraphics[width=\textwidth]{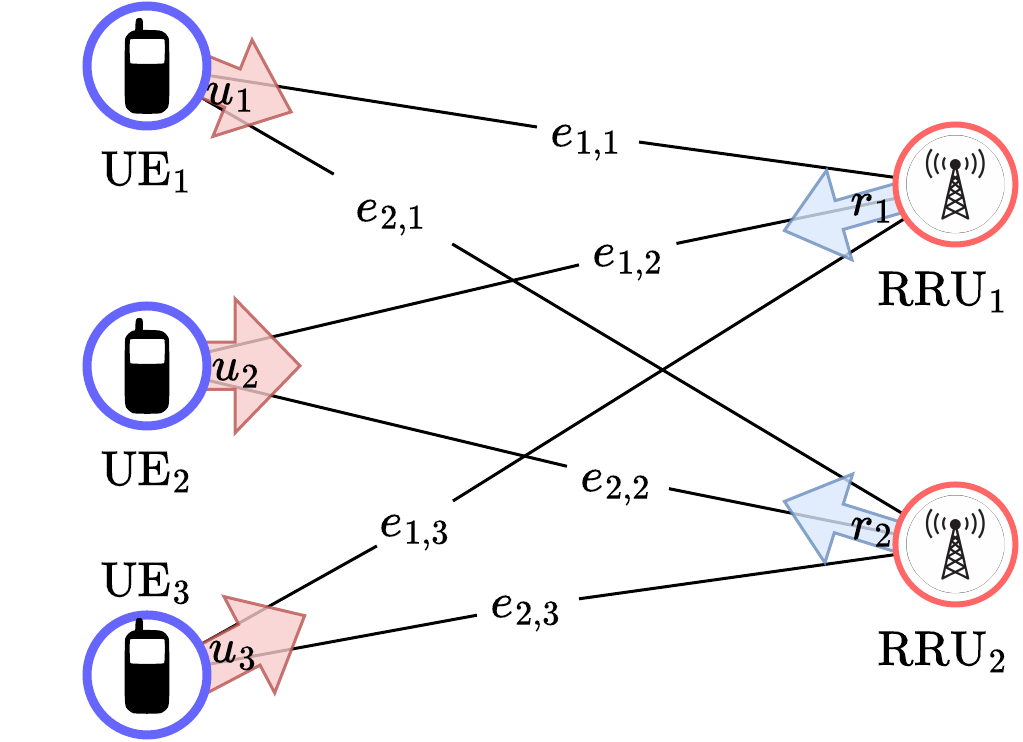}
    \caption{\textcolor{black}{The integrated sensing and communication graph for the system with $N=3$ UEs, and $K=2$ RRUs.}}
    \label{fig:graph_present}
    \end{minipage}\hfill
    \begin{minipage}{0.65\textwidth}
    \includegraphics[width=\textwidth]{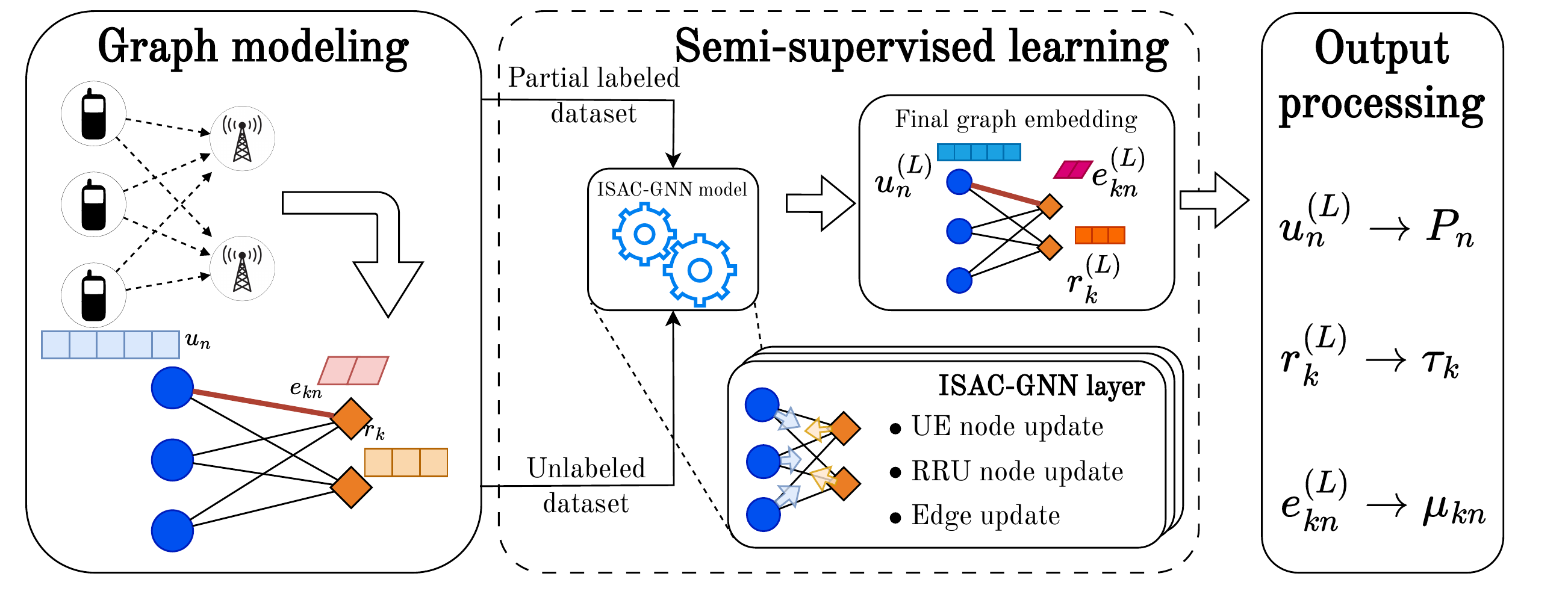}
    \caption{Architecture of ISAC-GNN model.}
    \label{fig:isac-gnn-architecture}
    \end{minipage}
    \vspace{-0.3cm}
\end{figure*}

In this subsection, we introduce ISAC-GNN, a GNN architecture designed to exploit the permutation equivariance property of graph-structured data. Fig.~\ref{fig:isac-gnn-architecture} presents the proposed ISAC-GNN, which comprises three key components: \textit{(i)} an initializing layer, \textit{(ii)} multiple ISAC-GNN layers, and \textit{(iii)} an output processing layer, each of which is detailed below.

\subsubsection{Initializing Layer}
This initializing layer prepares the node feature and edge feature for the ISCG as described in \ref{sec:graph_model}. The $n$-th UE node and the $k$-th RRU node are initialized as $u_{n}^{(0)} = [P^{\textrm{max}}_{n}, P_{\textrm{cir}},\kappa,\hat{x}_{n}, \hat{y}_{n}]$, and $r_{k}^{(0)} = [\tilde{x}_{k}, \tilde{y}_{k}, P_{\textrm{sens}}]$, respectively. \textcolor{black}{The edge connecting the $n$-th UE node and the $k$-th RRU node is initialized with attribute $e_{kn}^{(0)}[g_{kn}, \rho_{kn}]$.}

\subsubsection{ISAC-GNN Layers} After graph modeling the problem, we pass the graph structured data through $L$ ISAC-GNN layers. At the $(l)$-th layer, the UE nodes feature $u_{n}^{(l)}$, the RRU nodes feature $r_{k}^{(l)}$, and edges attribute $e_{kn}^{(l)}$ \textcolor{black}{are updated as follows}

\noindent$\bullet$ \textbf{UE node aggregation and update:}
\begin{align}   \label{gnn_ue} 
    \begin{aligned}
        m(v_{n}^{\textrm{u}})^{(l)} &= \underset{v_{k} \in\mathcal{V}^{\textrm{r}}}{\textrm{SUM}} \left\{ \Omega_{1}^{(l)} \left( r_{k}^{(l-1)}, e_{kn}^{(l-1)} \right) \right\}, \\
        u_{n}^{(l)} &=  \Omega_{2}^{(l)} \left( u_{n}^{(l-1)}, m(v_{n}^{\textrm{u}})^{(l)} \right).
    \end{aligned}
\end{align}
\noindent$\bullet$ \textbf{RRU node aggregation and update:}
\begin{align}\label{gnn_rru}
    \begin{aligned}
        m(v_{k}^{\textrm{r}})^{(l)} &= \underset{v_{n} \in\mathcal{V}^{\textrm{u}}}{\textrm{SUM}} \left\{ \Omega_{3}^{(l)} \left( u_{n}^{(l-1)}, e_{kn}^{(l-1)} \right) \right\}, \\
        r_{k}^{(l)} &=  \Omega_{4}^{(l)} \left( r_{k}^{(l-1)}, m(v_{k}^{\textrm{r}})^{(l)} \right).
    \end{aligned}
\end{align}
\noindent$\bullet$ \textbf{Edge update:}
\begin{align} \label{gnn_edge}
    e_{kn}^{(l)} &= \Omega_{5}^{(l)} \left( r_{k}^{(l)}, e_{kn}^{(l-1)}, u_{n}^{(l)} \right).
\end{align}
where $\Omega_{i}^{(l)}, i = [1,\dots,5]$ are the processing multi-layer perceptron (MLPs) at the $(l)$-th layer. 
The universal approximation theorem \cite{MLP_Apprx_1, MLP_Apprx_2} guarantees the expressiveness of ISAC-GNN, ensuring it can approximate any continuous function with minimal error. This allows ISAC-GNN to effectively capture complex graph-structured features in ISCG. Additionally, its scalability is ensured by the fact that node-wise processing in \eqref{gnn_ue} - \eqref{gnn_edge} remains independent of the network size, and the MLP output dimensions are fixed regardless of the number of UEs and RRUs. Thus, ISAC-GNN generalizes well across different system configurations while maintaining computational efficiency.
\subsubsection{Output Processing Layer} After $L$ ISAC-GNN layers, we post-process the final node feature and edge attribute to derive the uplink resource management policies. First, for the $n$-th UE, the transmit power is calculated based on the UE final node embedding $u_{n}^{(L)}$ as
\begin{align}
    P_n = \sigma \left(u_{n}^{(L)} \right) \times P^{\textrm{max}}_{n},
    \label{eq:power}
\end{align}
where $\sigma(\cdot)$ is the Sigmoid function. Therefore, the final power allocation will satisfy the limit power constraint \eqref{V1b}. 
Next, the activation status of the $k$-th RRU is determined as follows
\begin{align}
    \tau_{k} = \begin{cases}
        1, &\sigma\left(r_{k}^{(L)} \right) \geq 0.5, \\
        0, &\textrm{otherwise.}
    \end{cases}
\end{align}
An RRU remains active for sensing if $\sigma(r_{k}^{(L)}) \geq 0.5$; otherwise, it is deactivated.
Finally, the user association variable for the $n$-th UE is calculated as
\begin{align}
    \mu_{kn} = \begin{cases}
        1,\quad & \textrm{if }k = \underset{k'\in\mathcal{K}}{\textrm{argmax}}\frac{\textrm{exp}\Big(e_{k'n}^{(L)}\Big)}{\sum_{j \in \mathcal{K}} \textrm{exp}\Big(e_{jn}^{(L)}\Big) }, \\
        0, &\textrm{otherwise}.
    \end{cases}
    \label{eq:AP}
\end{align}
where $e_{kn}^{(L)}$ represents the final attribute of the edge between the $k$-th RRU and the $n$-th UE. This mechanism transforms the hidden states of all UE-RRU pairs into a scheduling probability distribution over the RRUs for each UE, ensuring compliance with constraint \eqref{V1c}. Consequently, each UE is assigned to the RRU with the highest probability, setting $\mu_{kn} = 1$ for the selected RRU and $\mu_{kn} = 0$ otherwise.

\subsection{Loss Function}
In this paper, we propose a semi-supervised learning framework to train ISAC-GNN, leveraging supervised learning for parameter initialization and unsupervised fine-tuning to enhance performance and avoid poor initialization or local optima.
The supervised loss $\mathcal{L}_{\textrm{sup}}$ is defined as
\begin{align}\label{eq:sup_loss}
        \begin{aligned}
        \mathcal{L}_{\textrm{sup}} &= \sum_{n=1}^{\mathcal{N}} \textrm{MSE}\left(P_{n}, P^{*}_{n} \right) + \sum_{k=1}^{\mathcal{K}} \textrm{BCE}\left(\tau_{k}, \tau^{*}_{k} \right) \\
        &+ \sum_{n=1}^{\mathcal{N}}\sum_{k=1}^{\mathcal{K}} \textrm{BCE}\left(\mu_{kn}, \mu^{*}_{kn}\right),
        \end{aligned}
\end{align}
where $P^{*}_{n}, \tau^{*}_{k},$ and $\mu^{*}_{kn}$ are the power allocation, RRU activation, and user association labels generated by the Algorithm \ref{AAlgorithm}, respectively. 
Here, continuous power variables are supervised using the mean square error (MSE), while discrete variables of RRU activation $\tau^{*}_{k}$ and user association $\mu^{*}_{kn}$ are supervised using binary cross-entropy (BCE). This mixed-variable formulation ensures that the ISAC-GNN is trained consistently across all resource management tasks.
 The unsupervised loss is defined as 
\begin{align}
    \mathcal{L}_{\textrm{unsup}} = - \eta(\boldsymbol{\Psi},\boldsymbol{\tau},\mathbf{p}) + \varsigma (\mathbf{b} - \kappa \mathbf{A}) \mathbf{p},
\end{align}
which seeks to maximize the EE over the considered ISAC system, and the penalty factor $\varsigma$ ensures that the solution satisfies the constraints in \eqref{P2:sense}. Then, the semi-supervised loss function, weighted by factor $\vartheta\in[0,1]$, is expressed as
\begin{align}\label{eq:semi_loss}
    \mathcal{L} = \vartheta \mathcal{L}_{\textrm{sup}} + (1-\vartheta) \mathcal{L}_{\textrm{unsup}}.
\end{align}
This semi-supervised approach combines the stability of supervised learning for initialization with the flexibility of unsupervised learning for fine-tuning, leading to better energy efficiency and overall training performance.

\subsection{Complexity Analysis}

In this section, we analyze the computational complexity of the proposed ISAC-GNN model. Following the approach in \cite{2024_SensJ_Giang_HGNN}, we divide the computational cost of GNN-based models into two main components: (i) the forward pass and (ii) the backpropagation. The complexity of each component is examined in the subsequent discussion.

\subsubsection{Forward Pass} The computation cost for each step in forward pass is divided into four distinct steps as following. For clarity, the superscript layer index $(l)$ is omitted.
\begin{itemize}
    \item \textbf{Message Generation:} As presented in \eqref{gnn_ue} and \eqref{gnn_rru}, a message is generated by passing an input $\mathbf{X}$ through a MLP $\Omega(\cdot)$, which can be generalized as $\varrho(\mathbf{X}\mathbf{W})$, where $\mathbf{W}$ represents the set of parameters of the MLP, and $\varrho(\cdot)$ is a non-linear activation function. Given that at the $l$-th ISAC-GNN layer, $\mathbf{X}$ has dimensions $(\vert \mathcal{V} \vert \times F_l)$ and $\mathbf{W}$ has dimensions $(F_{l} \times F_{l+1})$, the complexity of this step is $\bigO(\vert \mathcal{V} \vert \times F_{l} \times F_{l+1} + \vert \mathcal{V} \vert)$. Without loss of generality, assuming a constant embedding size $F$ across all layers, this simplifies to $\bigO(\vert \mathcal{V} \vert F^2 + \vert \mathcal{V} \vert)$.
    \item \textbf{Message Passing:} For each edge $e \in \mathcal{E}$, two messages of size $F$ are aggregated, resulting in a total cost of $\bigO(2\vert \mathcal{V} \vert)$.
    \item \textbf{Node Embedding Update:} Applying the second MLP $\Omega(\cdot)$ has the same computational complexity as the first step of $\bigO(\vert \mathcal{V} \vert F^2 + \vert \mathcal{V} \vert)$.
    \item \textbf{Edge Embedding Update:} Similarly, applying the third MLP $\Omega(\cdot)$ to edges yields the computational complexity of $\bigO(\vert \mathcal{E} \vert F^2 + \vert \mathcal{E} \vert)$.
\end{itemize}
Thus, the total computational cost for the forward pass over $L$ layers is
\begin{align} \label{eq:computation_complexity}
    \bigO(2L\vert \mathcal{V} \vert F^2  + 2L\vert \mathcal{V} \vert + 3L\vert \mathcal{E} \vert + L\vert \mathcal{E} \vert F^2).
\end{align}
\subsubsection{Backward Pass}
In reverse-mode automatic differentiation, the backward pass computes the gradients based on the computation graph from the forward pass. Consequently, the computational cost of the backward pass is generally the same as the forward pass, as given in  \eqref{eq:computation_complexity}. With the initial features maximum dimension of $F$, the total computational cost for the proposed $L$-layer ISAC-GNN over the ISCG with $N$ UEs and $K$ RRUs is thus
{\small
\begin{align}\label{eq:complex_GNN}
    \bigO\Bigg(2L\bigg((NK + 2N + 2K + 1)F^2  + 2(3NK + N + 4K + 1) \bigg)\Bigg).
\end{align}
}%

Compared to the model-based algorithm complexity in~\eqref{eq:complex_model}, which involves multiple nested iterative procedures and high computational overhead, the proposed ISAC-GNN exhibits significantly lower and more scalable computational complexity. Once trained, the GNN operates in linear time with respect to the number of UEs and RRUs, enabling fast and efficient deployment in large-scale ISAC systems.

\section{Theoretical Analysis of ISAC-GNN}\label{sec:Theoretical}

In this section, we theoretically evaluate the performance and generalization capabilities of the ISAC-GNN. We first compare its performance against a set of distributed computing algorithms. Building on this perspective, we then analyze how ISAC-GNN performs relative to the proposed Algorithm~\ref{AAlgorithm} for the EE maximization problem.

Prior studies have established equivalences between neural networks and optimization algorithms. For instance, \cite{sato2019approximation} demonstrated the equivalence of Spatial Graph Neural Networks and graph optimization, while \cite{Shen2021} analyzed Message Passing GNNs and the WMMSE \cite{WMMSE_base_paper} algorithm in power allocation problems. Building on these insights, this section compares the performance of ISAC-GNN with the proposed Algorithm~\ref{AAlgorithm} in Section \ref{sec:OptimiaztionApproach}.

\subsection{Performance of the ISAC-GNN and Distributed Computing Optimization}

First, we introduce the \textit{Distributed Broadcasting Computing Algorithm} (DBCA) \cite{hella2015weak}, a class of distributed computing algorithms that iteratively exchange messages in a multi-agent system. 
This class of algorithm covers various radio resource management techniques in wireless networks, such as the well-known WMMSE algorithm \cite{WMMSE_base_paper}.



A clear similarity exists between the message-passing mechanism in ISAC-GNN and the agent update process in DBCA. In each communication round, DBCA proceeds as follows
\textit{(i)} agents broadcast their internal states;
\textit{(ii)} receive messages from neighbors;
\textit{(iii)} process the received messages; and
\textit{(iv)} update their internal states.
Similarly, ISAC-GNN aggregates information from neighbors and updates node and edge representations at each layer. The formal equivalence is detailed in the following proposition.

\begin{proposition} \label{prop:DBCA-GNN} (Equivalence of ISAC-GNN and DBCA)
We denote DBCA$(L)$ as the branch of DBCA with $L$ communication rounds, and ISAC-GNN$(L)$ as the ISAC-GNN model with $L$ layers based on \eqref{gnn_ue}-\eqref{gnn_edge}. Then, the following results hold:
\begin{enumerate}
    \item For any ISAC-GNN$(L)$, there exists a distributed computing algorithm in DBCA$(L)$ that solves the same set of problems as ISAC-GNN$(L)$.
    \item For any algorithm in branch DBCA$(L)$, there exists an ISAC-GNN$(L)$ that solves the same set of problems as this algorithm.
\end{enumerate}
\end{proposition}
\begin{proof}
    Please refer to Appendix \ref{Proof:DBCA-GNN} for the detailed proof.
\end{proof}

This equivalence enables a theoretical performance analysis of ISAC-GNN by examining the behavior of DBCA. The proposition establishes that if no DBCA can solve a given radio resource management problem, then ISAC-GNN is also unable to solve it. Conversely, if a problem is well-suited for DBCA, ISAC-GNN can solve it with equal or competitive performance. In the next subsection, we apply this result to the EE maximization problem presented in Section \ref{sec:SystemModel}.

\subsection{Performance of the ISAC-GNN in energy efficiency maximization problem.}
In this subsection, the DBCA is used to analyze the performance and generalization capabilities of ISAC-GNN in the EE maximization problem. The comparison is based on the observation that, under certain conditions, the proposed Algorithm \ref{AAlgorithm} is categorized as a DBCA. This observation is formalized in the following lemma
\begin{lemma}\label{lemma:DBCA-AA}
    When the maximal number of neighbors is bounded by some constant, then the proposed Algorithm~\ref{AAlgorithm} with a fixed number of iterations is equivalent to a DBCA.
\end{lemma}
\begin{proof}
    Under the assumption of the lemma, it is shown that, at each update iteration of the proposed Algorithm \ref{AAlgorithm}, the DBCA can perform the same updates through message broadcasting and agent updating mechanisms. 
    The detailed proof is provided in Appendix \ref{Proof:DBCA-AA}.
\end{proof}
Lemma \ref{lemma:DBCA-AA} establishes that the AA algorithm operates as a DBCA under the specified conditions. Consequently, for the same training and testing problem sizes, there always exists an ISAC-GNN that achieves comparable performance to the AA algorithm in solving EE maximization problems.

\section{Numerical Results}\label{sec:Sim}

In this section, we analyze the performance of the proposed model-based and learning-based approaches across various system scenarios. To benchmark Algorithm \ref{AAlgorithm}, we introduce reference solutions for each sub-problem. Furthermore, we conduct extensive simulations to evaluate the convergence, scalability, and generalization capabilities of ISAC-GNN.

\subsection{Simulation Setups}

In all simulations, we consider a system with $K$ RRUs and $N$ UEs, along with a single target device. All devices are uniformly distributed within a two-dimensional coordinate system, constrained within a circular area of radius $r\mathrm{[m]}$, centered at $(r, r)\mathrm{[m]}$. \textcolor{black}{According to the key parameter indicators (KPIs) of 5G New Radio (NR) Release 17 \cite{3gpp_tr_38855},  a horizontal accuracy of meter-level is targeted for industrial IoT use cases. Thus, we select the localization accuracy as $\kappa=1 \mbox{ [m}^2\mbox{]}$ to reflect this standard-compliant localization requirement.}
The detailed system parameters are provided in Table \ref{Table1}.
\begin{table}[t] 
  \vspace{0.2cm}
  \renewcommand{\arraystretch}{1.25}
  \caption{Key Parameters for System Simulation}
  \centering
  \begin{tabular}{p{0.3\textwidth}|p{0.1\textwidth}}
      \toprule[1pt]\midrule[0.3pt] 
      Area radius $(r)$ & $200 \mbox{ [m]}$\\ 
      \midrule 
      Bandwidth $(B)$   & $100 \mbox{ [MHz]}$\\ 
      \midrule 
      Power spectral density of the thermal noise & $-174\mbox{ [dBm/Hz]}$ \\  
      \midrule
      UE Maximum transmit power ($P^{\textrm{max}}$) & $200 \mbox{ [mW]}$\\ 
      \midrule 
      UE Circuit power consumption ($P_{\textrm{cir}}$) & $100 \mbox{ [mW]}$\\ 
      \midrule 
      RRU Sensing power consumption ($P_{\textrm{sens}}$) & $50 \mbox{ [mW]}$\\ 
      \midrule 
      Sensing localization accuracy $\kappa$    &$1\mbox{ [m}^2\mbox{]}$\\
      \midrule
      Lagrange multiplier Step-size ($\varepsilon$) & $0.0001$\\ 
      \midrule 
      Algorithm stop condition ($\epsilon_1, \epsilon_2, \epsilon_3$) & $0.0001$\\ 
      \midrule 
      \bottomrule[1pt] 
  \end{tabular}
  \label{Table1}
  \vspace{-0.3cm}
\end{table}%

All channels follow an uncorrelated Rayleigh distribution, with large-scale fading given by $\iota_{kn} = \textrm{FSPL}_{kn} + \upsilon_{kn}$, where $\textrm{FSPL}_{kn} = -120.9 - 37.6 \log{10} d_{kn} \mbox{ [dB]}$ represents free-space path loss, and $\upsilon_{kn}$ follows a log-normal Gaussian distribution with a $7 \mbox{ [dB]}$ standard deviation. 
For the model-based approach, we evaluate $100,000$ realizations with $5,000$ random UE and $200$ RRU locations.
For learning-based models, we verify the generalization by training a single ISAC-GNN model across multiple $(N,K)$ settings and evaluating it on all test configurations without retraining. For each $(N,K)$ configuration we generate $2,000$ training and $500$ testing samples. 
For every realization, we run Algorithm~\Ref{AAlgorithm} to obtain the solution labels $(\mathbf{p}^*, \boldsymbol{\mu}^*, \boldsymbol{\tau}^*)$ for power allocation, user association, and RRU activation policies, respectively.
The MLPs in \eqref{gnn_ue} - \eqref{gnn_edge} use fully connected layers with ReLU (Rectified Linear Unit) activations. All models are train with $100$ epochs, learning rate of $10^{-4}$, and a batch size of $32$.
Regarding the semi-supervised loss in \eqref{eq:semi_loss}, we set $\vartheta=1$ for the first epoch and then gradually decrease it to $\vartheta=0$ by the $20$th epoch. For the remaining epochs, ISAC-GNN is trained in a fully unsupervised manner, directly maximizing the total system EE. This schedule provides a stable supervised warm start and then prioritizes the task objective.
All training and simulations were conducted on an Intel(R) Xeon(R) CPU~@~2.20 GHz.

\subsection{Model-based Approach Evaluation}

 The proposed scheme is the combination of the proposed fixed-point power allocation, RRU activation, and user association. To evaluate the performance of the AA algorithm in Algorithm \ref{AAlgorithm}, we compare it with several baseline methods. Specifically, each sub-problem is solved using the following benchmark approaches:
\begin{itemize}
    \item \textbf{Equal power allocation:} We use the golden search method to calculate the equal value to all UEs which can provide the highest EE to the system.
    \item \textbf{Nearest RRU selection:} The nearest RRU is assigned to each UE for the best communication channel.
    \item \textbf{Full RRU activation:} We activate all RRUs.
\end{itemize}
To ensure a comprehensive evaluation, the proposed algorithm is tested in combination with these benchmark approaches, resulting in \textit{eight} distinct models.
For brevity, each model is denoted by a short descriptive label instead of the full name:  
\begin{itemize}
    \item \textbf{P}: the proposed power allocation; and \textbf{Equal}: equal power allocation,
    \item \textbf{U}: the proposed user association; and \textbf{Near}: nearest RRU selection,
    \item \textbf{R}: the proposed RRU activation; and \textbf{Full}: full RRU activation.
\end{itemize}
For example, the combination of proposed power allocation, full RRU activation, and nearest RRU selection is denoted as \textit{P-Near-Full}, while the fully benchmark-based model is represented as $\textit{Equal-Near-Full}$. This notation is used consistently in all figures for clarity.
Moreover, we employ another baseline as described in \cite{2025_TVT_Tung_HetGNN}, namely \textbf{Decoupled Method}. This method tackles the joint optimization problem in a block-separable approach, where the communication-only problem is first obtained,
\begin{subequations}
    \begin{align}
        \underset{\substack{ \boldsymbol{\Psi}, \mathbf{p}, \boldsymbol{\tau} }}{\mathrm{maximize} } &\quad  \eta(\boldsymbol{\Psi},\boldsymbol{\tau},\mathbf{p}) \\
        \mbox{s.t.}\quad&\quad \eqref{V1b}, \eqref{V1RRU}, \eqref{V1c} .\nonumber
    \end{align}
\end{subequations} 
Then, to satisfy the CRLB constraint \eqref{V1d}, the power allocation is scaled by $\delta = \textrm{max}\left( \tfrac{\mathbf{b}^T \mathbf{p}}{\mathbf{p}^T \mathbf{A} \mathbf{p}}\right)$. This baseline provides a fair and widely-recognized benchmark to compare against our proposed scheme.

\begin{figure}[t]
    \centering
    \includegraphics[width=0.9\columnwidth]{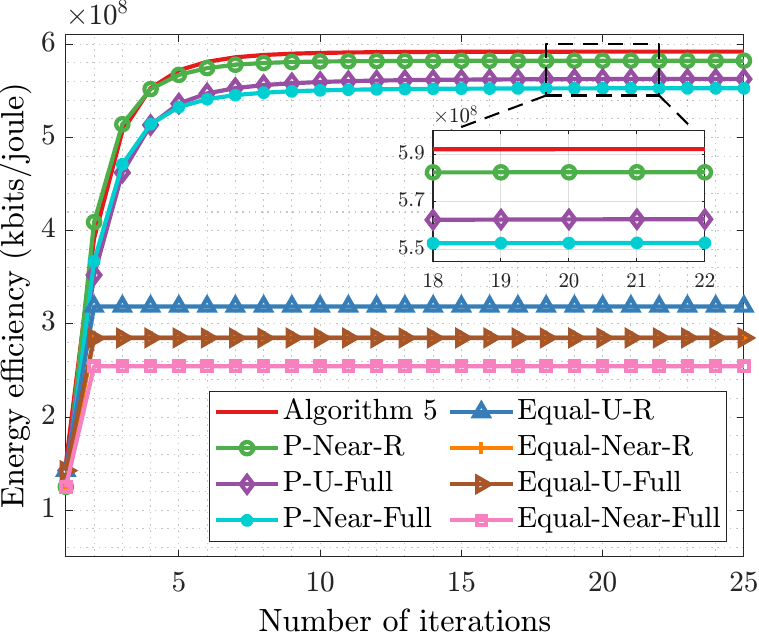}
    \caption{The EE versus number of iterations in Algorithm \ref{AAlgorithm} \textcolor{black}{compared} with different resource management with $K=8, N=24$.}
    \label{fig:converge_AA}
    \vspace{-0.2cm}
\end{figure}

\begin{figure}[t]
\vspace{1em}
    \centering
    \includegraphics[width=0.9\columnwidth]{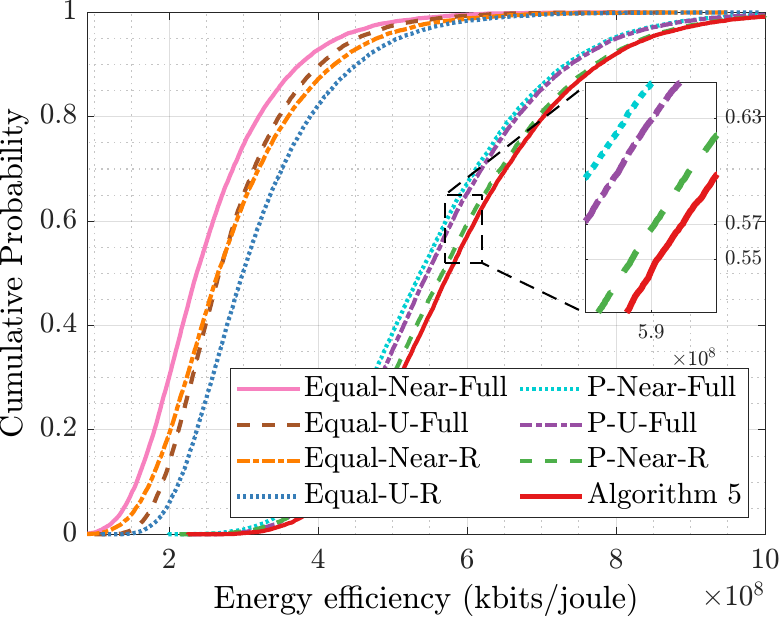}
    \caption{The CDF of maximum EE at the end iteration in Algorithm \ref{AAlgorithm} \textcolor{black}{compared} with different resource management with $K=8, N=24$.}
    \label{fig:CDF_vs_bench}
    \vspace{-0.1cm}
\end{figure}

Fig.~\ref{fig:converge_AA} illustrates the EE convergence of various model-based methods over iterations for a system with $K=8$ RRUs and $N=24$ UEs. The proposed Algorithm~\ref{AAlgorithm} consistently outperforms all benchmarks, achieving a peak EE of approximately $5.9 \times 10^8\mbox{ [kbits/Joule]}$.
The significant impact of the proposed fixed-point power allocation is evident, as schemes using equal power allocation, such as `Equal-U-R' and `Equal-Near-R', only reach $3.2 \times 10^8$ and $2.8 \times 10^8$ kbits/Joule, respectively. Additionally, the proposed RRU activation enhances system EE by $6\%$, as observed in the gap between Algorithm~\ref{AAlgorithm} and `P-Near-R' schemes.
Fig.~\ref{fig:CDF_vs_bench} presents the cumulative distribution function (CDF) of the achieved EE across different models in the same system setup, further emphasizing the performance gap of Algorithm\ref{AAlgorithm} compared to benchmark methods.
About $45\%$ of the $100,000$ realizations surpass $5.9 \times 10^8\mbox{ [kbits/Joule]}$ with the full proposed scheme, whereas only $37\%$ of `P-U-Full' realizations achieve this due to the absence of optimized RRU activation. Notably, without the proposed power allocation method, no scheme attains this EE level.

\begin{figure}[t]
    \centering
    \includegraphics[width=0.9\columnwidth]{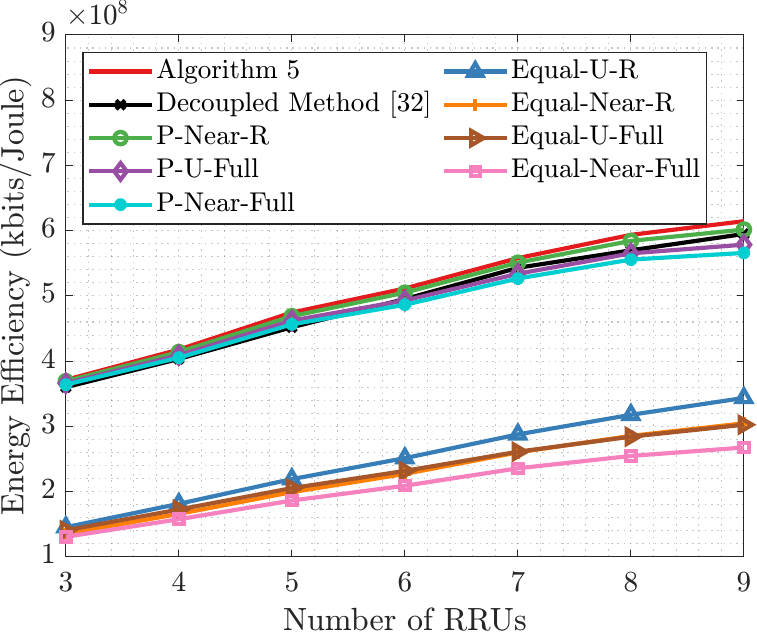}
    \caption{The maximum EE versus the number of RRUs in Algorithm \ref{AAlgorithm} \textcolor{black}{compared} with different resource management with $N=24, K=3:9$.}
    \label{fig:EE_vs_RRUs}
    \vspace{-0.1cm}
\end{figure}

\begin{figure}[t]
    \centering
    \includegraphics[width=0.9\columnwidth]{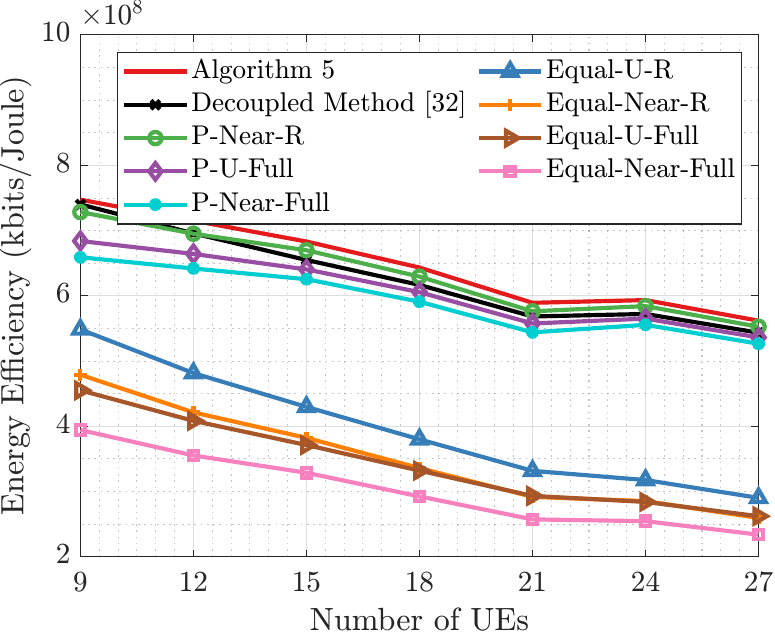}
    \caption{The maximum EE versus the number of UEs in Algorithm \ref{AAlgorithm} \textcolor{black}{compared} with different resource management with $N=9:3:27, K=8$.}
    \label{fig:EE_vs_UEs}
    \vspace{-0.1cm}
\end{figure}

Fig.~\ref{fig:EE_vs_RRUs} and \ref{fig:EE_vs_UEs} compare the maximum EE as a function of the number of RRUs and UEs in the ISAC system. In Fig.~\ref{fig:EE_vs_RRUs}, EE increases with more RRUs, as UEs benefit from better channel conditions and lower interference.
The proposed algorithm consistently outperforms all benchmarks. While the fully benchmark-based method achieves only $35\%$ of the EE at $K=3$ and $42\%$ at $K=9$ compared to the proposed approach, the Decoupled Method achieves better performance than simple baselines but still falls short of our proposed scheme. This is due to the scaling process, increasing the total power consumption without considering the overall EE.
By constrast, Fig.~\ref{fig:EE_vs_UEs} shows that increasing UEs while keeping RRUs fixed heightens interference, raising power consumption and lowering EE. Despite EE dropping from $7.4 \times 10^8$ to $5.6 \times 10^8$ kbits/Joule as $N$ grows from $9$ to $27$, Algorithm~\ref{AAlgorithm} consistently outperforms all benchmarks, including the Decoupled Method. In the most congested scenario of $N=27$, the benchmark-based method achieves only $41\%$ of the EE provided by the proposed approach.
Compared with the Decoupled Method, Algorithm~\ref{AAlgorithm} avoids the post-scaling power inflation incurred when enforcing the CRLB after a communication-only solve; by optimizing power and the sensing CRLB constraint jointly, it meets the sensing requirement with lower transmit power yielding higher EE.

\subsection{Learning-based Approach Evaluation}
\begin{figure}[t]
    \centering
    \includegraphics[width=0.9\columnwidth]{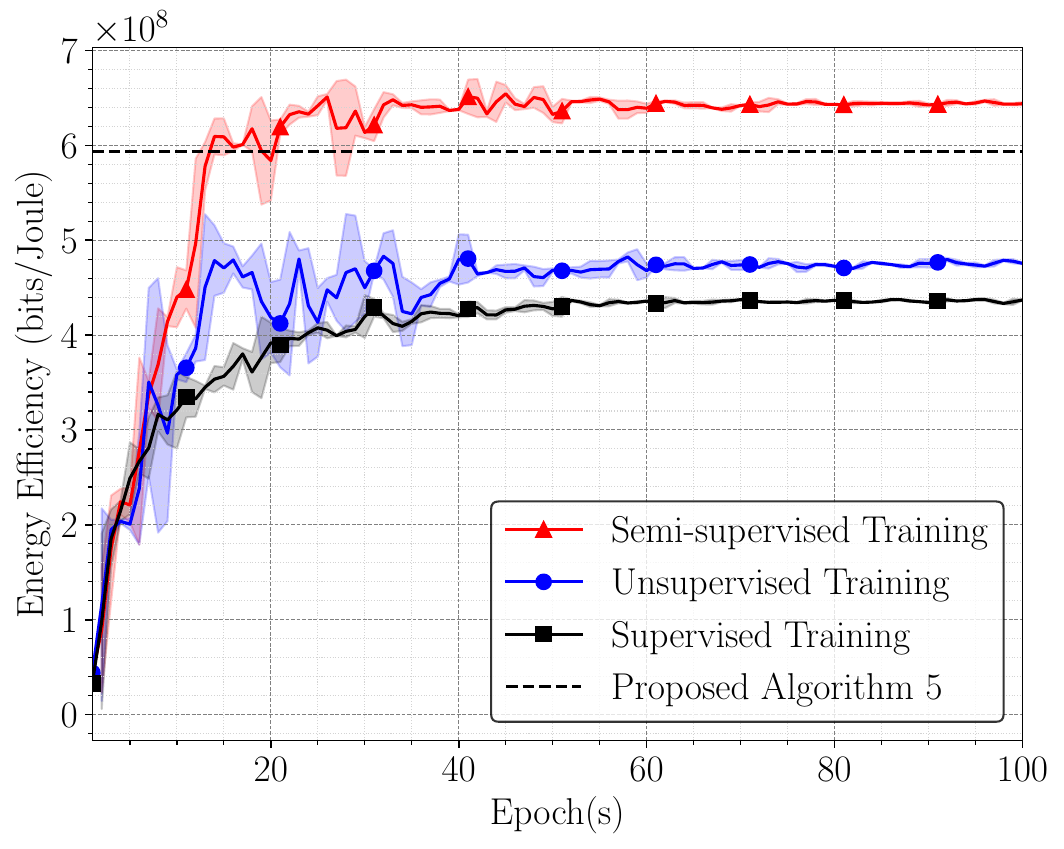}
    \caption{Training performance versus epochs for different settings: (i) semi-supervised, (ii) unsupervised, and (iii) supervised.}
    \label{fig:multi_training}
\end{figure}

\begin{table*}[t]
\renewcommand{\arraystretch}{2} 
\setlength{\tabcolsep}{0.5em} 
\caption{Performance comparison between learning-based, and model-based approaches with different system scenarios.}
\label{tab:Perf_comp}
\begin{minipage}{0.48\linewidth}
\centering
\resizebox{\columnwidth}{!}{
\begin{tabular}{@{\hspace{1em}}cccccc @{\hspace{1em}}}
\toprule \midrule
\multirow{2.5}{*}{\makecell{Number of \\ RRUs}} & \multirow{2.5}{*}{\makecell{Number of \\ UEs }}       & \multicolumn{2}{c}{Energy Efficiency $\times10^8 \mbox{[bits/Joule]}$} & \multicolumn{2}{c}{Run-time $\mbox{[ms]}$}   \\ 
\cmidrule(r){3-4} \cmidrule(lr){5-6}
                                                        && ISAC-GNN                                     &Algorithm~\ref{AAlgorithm}                             & ISAC-GNN       &Algorithm~\ref{AAlgorithm}\\ \hline
$K=8$     & $N=9$                       & \makecell{ $7.46$ \\ $99.78\%$ }              & \makecell{ $7.48$ \\ $100.00\%$ }             & $19.39$        & $80.23$       \\
$K=8$     & $N=12$                      & \makecell{ $7.26$ \\ $101.53\%$ }             & \makecell{ $7.15$ \\ $100.00\%$ }             & $19.56$        & $144.25$   \\
$K=8$     & $N=15$                      & \makecell{ $7.25$ \\ $104.91\%$ }             & \makecell{ $6.91$ \\ $100.00\%$ }             & $19.05$        & $310.61$   \\
$K=8$     & $N=18$                      & \makecell{ $6.79$ \\ $106.26\%$ }             & \makecell{ $6.39$ \\ $100.00\%$ }             & $19.67$        & $464.68$   \\
$K=8$     & $N=21$                      & \makecell{ $6.30$ \\ $106.19\%$ }             & \makecell{ $5.93$ \\ $100.00\%$ }             & $19.66$        & $575.52$   \\
$K=8$     & $N=24$                      & \makecell{ $6.30$ \\ $106.21\%$ }             & \makecell{ $5.93$ \\ $100.00\%$ }             & $19.60$        & $923.83$   \\
$K=8$     & $N=27$                      & \makecell{ $5.82$ \\ $103.23\%$ }             & \makecell{ $5.63$ \\ $100.00\%$ }             & $19.80$        & $1332.82$  \\ \midrule
\bottomrule     
\end{tabular}
}
\end{minipage}
\begin{minipage}{0.48\linewidth}
\resizebox{\columnwidth}{!}{
\begin{tabular}{@{\hspace{1em}}cccccc @{\hspace{1em}}}
\toprule \midrule
\multirow{2.5}{*}{\makecell{Number of \\ RRUs}} & \multirow{2.5}{*}{\makecell{Number of \\ UEs }}       & \multicolumn{2}{c}{Energy Efficiency $\times10^8 \mbox{[bits/Joule]}$} & \multicolumn{2}{c}{Run-time $\mbox{[ms]}$}   \\ 
\cmidrule(r){3-4} \cmidrule(lr){5-6}
                                                        && ISAC-GNN                                     &Algorithm~\ref{AAlgorithm}                             & ISAC-GNN       &Algorithm~\ref{AAlgorithm}\\ \hline

$K=3$     & $N=24$                      & \makecell{ $3.72$ \\ $100.61\%$ }             & \makecell{ $3.70$ \\ $100.00\%$ }             & $18.85$        & $434.52$      \\  
$K=4$     & $N=24$                      & \makecell{ $4.31$ \\ $102.27\%$ }             & \makecell{ $4.22$ \\ $100.00\%$ }             & $19.84$        & $621.25$      \\  
$K=5$     & $N=24$                      & \makecell{ $4.89$ \\ $103.84\%$ }             & \makecell{ $4.71$ \\ $100.00\%$ }             & $19.01$        & $660.07$      \\  
$K=6$     & $N=24$                      & \makecell{ $5.32$ \\ $104.90\%$ }             & \makecell{ $5.07$ \\ $100.00\%$ }             & $19.37$        & $811.29$      \\  
$K=7$     & $N=24$                      & \makecell{ $5.83$ \\ $105.43\%$ }             & \makecell{ $5.53$ \\ $100.00\%$ }             & $19.53$        & $796.72$      \\  
$K=9$     & $N=24$                      & \makecell{ $6.51$ \\ $104.95\%$ }             & \makecell{ $6.20$ \\ $100.00\%$ }             & $18.76$        & $933.98$      \\  
    &                   & \makecell{ \phantom{zz} \\\phantom{zz} }              &       &        &       \\ \midrule 
\bottomrule     
\end{tabular}
}
\end{minipage}
\vspace*{-0.25cm}
\end{table*}
Fig.~\ref{fig:multi_training} compares the training performance of ISAC-GNN under three settings: semi-supervised, unsupervised, and supervised, compared with the proposed Algorithm~\ref{AAlgorithm}. The shaded regions indicate variability, where narrower areas reflect greater stability.
Supervised training is highly stable, as shown by its narrow shaded region. However, it relies on labeled data, limiting generalization and failing to optimize the objective reward directly. Unsupervised training, in contrast, optimizes the reward but exhibits greater variability due to sensitivity to parameter initialization and local optima.
The proposed semi-supervised approach combines the strengths of both. It initially integrates supervised and unsupervised learning before fully transitioning to unsupervised training. This strategy ensures stable parameter initialization while effectively optimizing the objective, leading to the highest final performance with moderate variability.

Table~\ref{tab:Perf_comp} compares the performance of the learning-based model-based approaches to solve the EE maximization problem \eqref{Problemv1} by exploiting the ISAC-GNN in Section \ref{sec:ISAC-GNN} and Algorithm \ref{AAlgorithm} in \ref{sec:OptimiaztionApproach}, respectively. \textcolor{black}{In this simulation, we train the ISAC-GNN model with various numbers of UEs and RRUs with $N\in\{9,12,15,18,21,24,27\}$ and $K\in\{3,4,5,6,7,8,9\}$ and test the model with every scenario. }
Regarding the EE, \textcolor{black}{the ISAC-GNN outperforms the model-based approach in every scenario}. On average, the learning-based approach can provide $4\%$ better EE than the model-based Algorithm \ref{AAlgorithm}. It should be noticed that, the ISAC-GNN demonstrate the great scalability, we only train one model, and it can be applied to several system setups. Regarding running time, the learning-based model can predict the resource policies in the order of milliseconds, and this running time does not scale with the system size, \textcolor{black}{stabilizing at about $20 \mbox{ [ms]}$}. In contrast, Algorithm \ref{AAlgorithm} has much higher time consumption, which scales up to $67$ times in the scenario of $K=8$ RRUs and $N=27$ UEs. Importantly, the model-based approach does not scale well with the system size, \textcolor{black}{when we increase $N$ threefold from $N=9$ to $N=27$ UEs, the running time increases by almost $16\times$ times.}

\begin{figure}[t]
    \centering
    \includegraphics[width=0.9\linewidth]{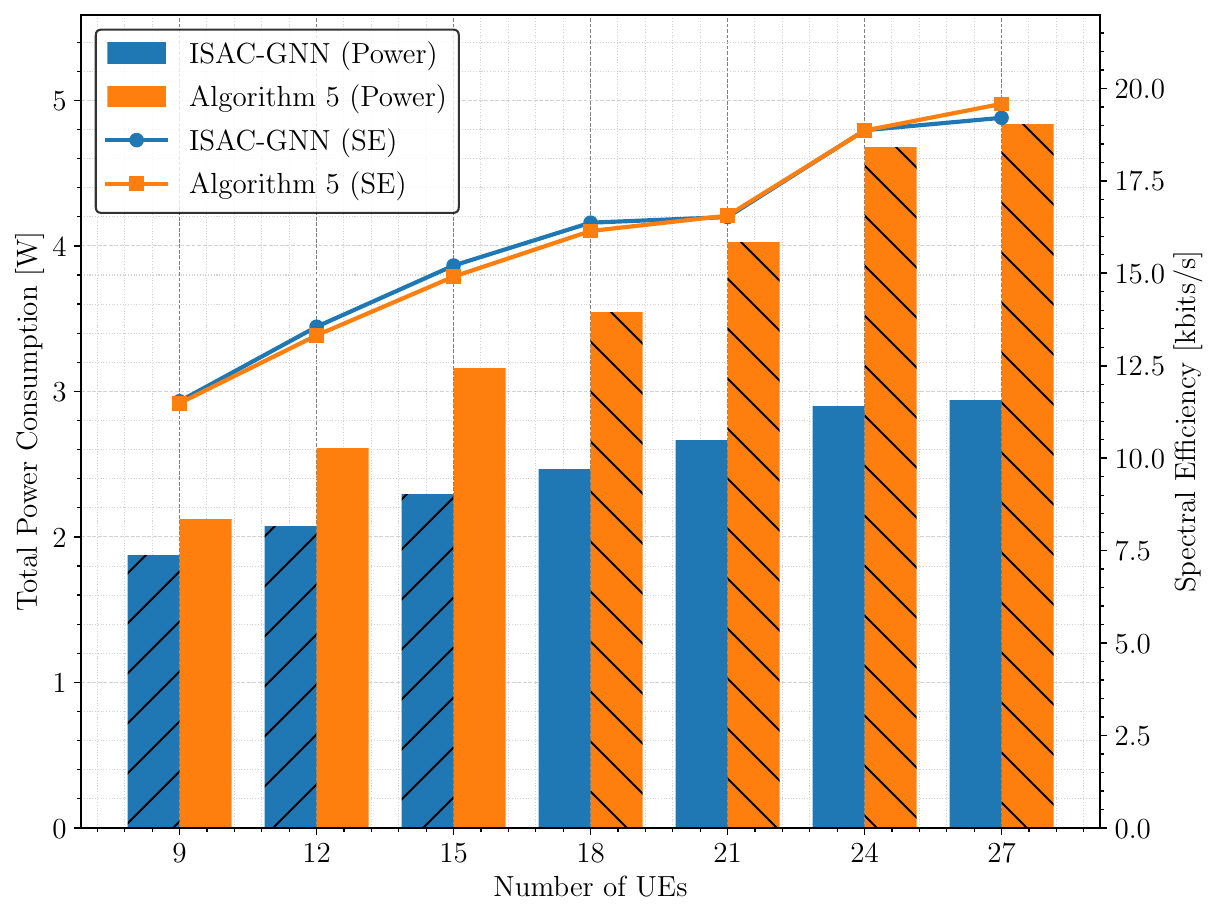}
    \caption{Total transmit power and spectral efficiency versus number of UEs.}
    \label{fig:power_SE}
\end{figure}
    
\textcolor{black}{Fig.~\ref{fig:power_SE} shows the total system power consumption}, shown as bars on the left axis, and the spectral efficiency (SE), shown as lines on the right axis, as the number of UEs increases.
In the bar chart, as the number of UEs increases, ISAC-GNN consistently requires less total power than Algorithm~\ref{AAlgorithm}. This is due to the joint optimization of UE transmit powers and RRU activation in ISAC-GNN, which eliminates redundant sensing and lowers interference. In particular, at $N=27$ UEs, ISAC-GNN consumes about $2.7 \mbox{ [W]}$ versus $3.5\mbox{ [W]}$ for Algorithm 5, a reduction of approximately $23\%$.
The line chart reports the achieved SE in the same network configurations. It can be seen that, the SE of the two methods are nearly identical over the entire range. Consequently, lower power consumption directly leads to higher total EE for the proposed ISAC-GNN. These results demonstrate that ISAC-GNN achieves more efficient power allocation than the model-based Algorithm~\ref{AAlgorithm} while preserving essentially the same SE, thereby yielding higher overall EE.

\begin{table}[t]
    \renewcommand{\arraystretch}{1.5} 
    \centering
    \caption{Scalability of ISAC-GNN with large-scale system.}
    \resizebox{\columnwidth}{!}{
    \begin{tabular}{@{\hspace{1em}}ccccccc@{\hspace{1em}}}
    \toprule \midrule
    
    \multirow{2.5}{*}{\makecell{Number of \\RRU}} &
    \multirow{2.5}{*}{\makecell{Number of \\ UE}} &
    \multicolumn{2}{c}{\makecell{Energy Efficiency\\$\times10^8 \mbox{[bits/Joule]}$}} &
    \multicolumn{2}{c}{Run-time $\mbox{[s]}$} \\
    \cmidrule(lr){3-4} \cmidrule(lr){5-6}
    &  & 
    ISAC-GNN & Algorithm~\ref{AAlgorithm} & 
    ISAC-GNN & Algorithm~\ref{AAlgorithm} \\
    \hline
    $K=40$      & $N=80$    & $7.25$    & $6.99$    & $0.019$   & $42.77$  \\
    $K=40$      & $N=120$   & $7.09$    & $6.84$   & $0.020$   & $59.39$  \\
    $K=60$      & $N=120$   & $7.45$    & $7.18$    & $0.019$   & $77.53$  \\
    $K=60$      & $N=160$   & $7.23$    & $6.89$   & $0.019$   & $154.78$  \\
    $K=80$      & $N=160$   & $7.54$    & $7.05$    & $0.020$   & $205.40$  \\
    $K=80$      & $N=200$   & $7.12$    & $6.85$   & $0.019$   & $420.35$  \\
    $K=100$     & $N=200$   & $7.32$    & $7.12$    & $0.020$   & $702.47$  \\
    \bottomrule 
    \end{tabular}
    }
    \label{tab:large_scale}
\end{table}

\begin{table}[t]
    \renewcommand{\arraystretch}{1.5} 
    \centering
    \caption{Performance of ISAC-GNN under different network densities with $N=24$, $K=8$.}
    \resizebox{\columnwidth}{!}{
    \begin{tabular}{@{\hspace{1em}}cccccccc@{\hspace{1em}}}
    \toprule \midrule
    \multirow{2.5}{*}{\makecell{Radius \\ \mbox{[m]}}} &
    \multicolumn{2}{c}{Density \mbox{[devices/km$^2$]}} & 
    \multicolumn{2}{c}{\makecell{Energy Efficiency\\$\times10^8 \mbox{[bits/Joule]}$}} & 
    \multicolumn{2}{c}{Run-time $\mbox{[ms]}$} \\
    \cmidrule(lr){2-3} \cmidrule(lr){4-5} \cmidrule(lr){6-7}
    &
    UE & RRU & 
    ISAC-GNN & Algorithm~\ref{AAlgorithm} & 
    ISAC-GNN & Algorithm~\ref{AAlgorithm} \\
    \hline
    $50$  & $3059$ & $1019$    & $9.01$ & $8.38$ & $19.53$ & $202.48$  \\
    $100$  & $764$ & $255$    & $7.68$ & $7.36$ & $19.98$ & $668.62$  \\
    $200$  & $191$ & $64$    & $6.22$ & $5.92$ & $18.62$ & $941.47$  \\
    $400$  & $48$ & $16$    & $5.12$ & $4.87$ & $19.24$ & $2130.60$  \\
    \bottomrule 
    \end{tabular}
    }
    \label{tab:density}
\end{table}

Table~\ref{tab:large_scale} and Table \ref{tab:density} demonstrate the generalization of the proposed ISAC-GNN to large-scale and various density system networks. Table~\ref{tab:large_scale} shows that ISAC-GNN maintains both accuracy and near-constant inference time as the network scales. In particular, when the system up scales from size of $(K,N)~=~(40, 80)$ to $(100,200)$, the proposed ISAC-GNN improves the EE by up to $7.5\%$ over the model-based Algorithm~\ref{AAlgorithm}. \textcolor{black}{While the latter run-time rises to more than $700 \mbox{ [s]}$, ISAC-GNN model inference time remains nearly constant, reducing runtime by $99.9\%$.} This is due to much lower inference-time complexity of the proposed ISAC-GNN w.r.t the system size. In Table~\ref{tab:density}, we fix the number of UEs and RRUs at $K~=~8, N~=~24$ and vary the coverage radius. Algorithm~\ref{AAlgorithm}  exhibits increasing run-time for sparser networks, as poorer overall CSI tightens constraints and requires more iterations to converge. By contrast, ISAC-GNN executes a single forward pass whose cost is essentially determined by graph size and layer widths, leading to nearly constant inference time throughout. 
These results substantiate the real-time efficiency and scalability of ISAC-GNN across large-scale and heterogeneous dense networks, while consistently outperforming the model-based approach.

\begin{figure}[t]
    \centering
    \includegraphics[width=0.9\columnwidth]{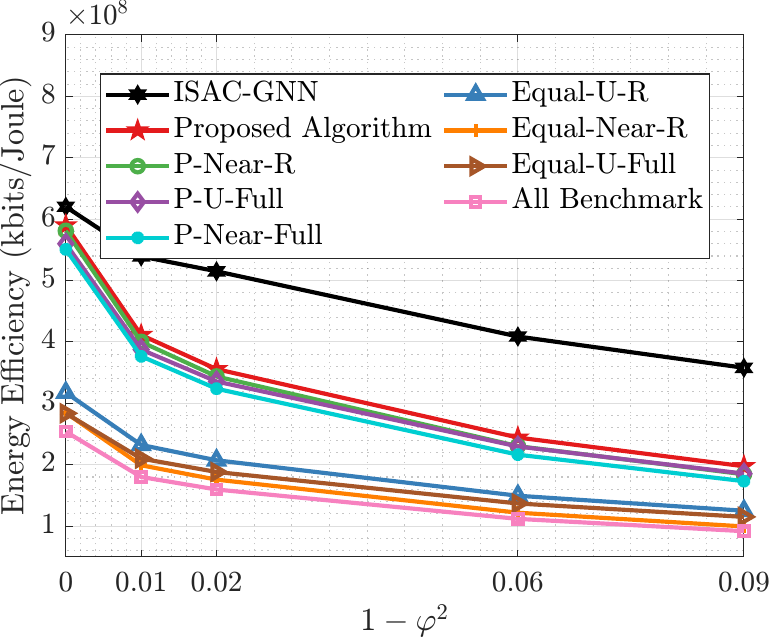}
    \caption{The achievable EE when the CSI is imperfect with $K=8, N=24$.}
    \label{fig:EE_imperfect_CSI}
    \vspace{-0.1cm}
\end{figure}
We evaluate the proposed ISAC-GNN under channel-estimation errors. For each UE-RRU link, the estimated channel is modeled as
\begin{equation}
\hat g_{kn} = \varphi g_{kn} , %
\end{equation}
where $g_{kn}$ is the true channel gain. Then the channel estimation error is given as $\sqrt{1-\varphi^{2}}\chi_{kn}$, where $\chi_{kn}\sim \mathcal{CN}(0,\bar{\iota}_{kn})$ models the estimation error with linear large‑scale power $\bar{\iota}_{kn} = 10^{\iota_{kn}/10}$, and $\varphi\in[0,1]$ is a channel-quality factor i.e.,$\varphi=1$ denotes perfect CSI.
Accordingly, we can calculate the SINR of $n$-th UE as
\resizebox{\linewidth}{!}{
$\hat{\gamma}_{n}=  \frac{\sum_{k\in\mathcal{K}}\varphi^{2}\vert\mu_{kn}g_{kn} \vert^{2}P_{n}}{\sum_{n^{\prime}\in \{\mathcal{N}\setminus n\}}\sum_{k\in\mathcal{K}}\vert\mu_{kn'}g_{kn} \vert^{2}P_{n'} + \sum_{k\in\mathcal{K}}\mu_{kn}(1-\varphi^{2})\bar{\iota}_{kn} +\sigma^2}. $
}
Apart from this modified SINR expression, all other procedures remain the same as in the perfect-CSI case. ISAC-GNN is trained on perfect-CSI data and tested at $\varphi\in\{1.0,0.995,0.99,0.97,0.95\}$. 
Fig.~\ref{fig:EE_imperfect_CSI} demonstrates achievable EE versus channel-quality factor under imperfect CSI. \textcolor{black}{It can be observed that EE of all schemes decreases as the channel-quality factor decreases, i.e., error factor $\sqrt{1-\varphi^2}$ increases.} This is owing to the fact that a higher error factor leads to lower received signal strength and higher interference, lowering the total achievable rate.
Among the model-based methods, all methods employing the proposed power allocation decrease more rapidly, compared to equal power allocation. This is due to the SINR is impacted by the lower signal strength and increased interference due to channel errors. By contrast, the ISAC-GNN still experiences performance degradation, but it is less severe than the proposed method in Algorithm~\ref{AAlgorithm}.  Notably, the gap between ISAC-GNN and other baselines widens as the CSI quality worsens.
Even when the error factor $1-\varphi^2$ is $0.09$, the GNN-based model maintains an EE of $3.5\times 10^8 \mbox{[bits/Joule]}$, which is still significantly higher than all model-based methods.
This indicates excellent generalization and robustness to imperfect CSI of the proposed ISAC-GNN.

We recognize that although our proposed ISAC-GNN model demonstrates efficiency and strong scalability, further enhancements should be considered for practical deployment under imperfect CSI. In particular, robustness can be strengthened by augmenting the training data with noisy CSI realizations or by combining GNN inference with online refinement techniques based on statistical error models or lightweight model-based updates. Such enhancements will further ensure the robustness and reliable deployment of ISAC-GNN in practical ISAC C-RAN systems.

\section{Conclusions}\label{sec:conclusion}
In this paper, we tackled the energy efficiency (EE) maximization problem in an uplink ISAC C-RAN system by jointly optimizing power allocation, user association, and selecting the best sensing RRUs while meeting the sensing accuracy requirement. To address the non-convexity of the problem, we proposed an alternative optimization algorithm, \textcolor{black}{which achieved approximately three times higher EE than the fully benchmark-based scheme.}
To enable real-time implementation, we introduced ISAC-GNN, a learning-based architecture that jointly optimizes all resource allocation variables. Leveraging semi-supervised training, ISAC-GNN achieves superior performance, providing an average EE gain of $5\%$ over the model-based approach while significantly reducing computational complexity, up to $64$ times faster in large-scale scenarios. Notably, ISAC-GNN demonstrates strong scalability, as a single trained model generalizes effectively across various system configurations with different numbers of UEs and RRUs.
These results underscore the potential of ISAC-GNN as a scalable and efficient solution for resource management in passive ISAC uplink C-RAN systems.

\appendices
\section{Proof of Lemma \ref{LogLemma}} \label{ProofLemmaLogApproximation}
We denote the optimal solution to the problem in \eqref{ProblemP_Algorithm} at the $i$-th iteration of Algorithm \ref{LogApproximationAlgorithm} as $\mathbf{p}^{(i)^*}$. Then, we have 

\begin{equation}
    \begin{aligned}
        \eta(\mathbf{p}^{(i-1)^*}) &\stackrel{(a)}{=} \tilde{\eta}(\mathbf{p}^{(i-1)^*},\boldsymbol{\alpha}(\mathbf{p}^{(i-1)^*}),\boldsymbol{\beta}(\mathbf{p}^{(i-1)^*})\\ &\stackrel{(b)}{\leq} \tilde{\eta}(\mathbf{p}^{(i)^*},\boldsymbol{\alpha}(\mathbf{p}^{(i-1)^*}),\boldsymbol{\beta}(\mathbf{p}^{(i-1)^*})\\
        & \stackrel{(c)}{\leq} \eta(\mathbf{p}^{(i)^*}) \stackrel{(a)}{=} \tilde{\eta}(\mathbf{p}^{(i)^*},\boldsymbol{\alpha}(\mathbf{p}^{(i)^*}),\boldsymbol{\beta}(\mathbf{p}^{(i)^*})).
    \end{aligned}
\end{equation}
The equality $(a)$ comes directly from \eqref{LowerboundInitial}. Next, because $\mathbf{p}^{(i)^*}$ is the optimal solution to the problem in \eqref{ProblemP_Algorithm}, and thus $(b)$ is always satisfied. From \cite{Singh2015}, {we have $\log_2(1+z)\geq \alpha \log(z) + \beta$ for any $\alpha$ and $\beta$}. Therefore, 
\begin{align}
    \begin{aligned}
        \log_2&(1+\gamma_{n}(\mathbf{p}^{(i)^*})) \\
        &\geq \alpha_{n}(\mathbf{p}^{(i-1)^*}) \log_2(\gamma_{n}(\textrm{\textbf{p}}^{(i)^*})) + \beta_{n}(\mathbf{p}^{(i-1)^*}).
    \end{aligned}
\end{align} 
Consequently, 
$\eta(\mathbf{p}^{(i)^*})\geq \tilde{\eta}(\mathbf{p}^{(i)^*},\boldsymbol{\alpha}(\mathbf{p}^{(i-1)^*}),\boldsymbol{\beta}(\mathbf{p}^{(i-1)^*}))$. Thus, the inequality $(c)$ is satisfied. In addition, $\eta(\mathbf{p})$ is bounded above due to the limited power budget. As a result, solving the problem \eqref{ProblemP0} using Algorithm \ref{LogApproximationAlgorithm}, the value of $\eta(\mathbf{p})$ always increases and converges. The proof is completed.
\section{Proof of Lemma \ref{SIFlemma}} \label{ProofLemma3}
\begin{table*}[b]
\vspace{-0.2cm}
\hrule
\vspace{0.2cm}
    \begin{equation}\label{DerivativeIF}
        \frac{\partial I_n(\mathbf{p})}{\partial P_n}=\frac{B\alpha_{n}\log_{2}(\textrm{e})\displaystyle\sum_{j\in \mathcal{N}\setminus \{n\}}B\alpha_{j}\frac{ \vert\boldsymbol{\mu}_j^T\boldsymbol{g}_{n}\vert^4}{\ln (2) \left( \displaystyle\sum_{n' \in \mathcal{N}\setminus \{j\}} \vert\boldsymbol{\mu}_{j}^T \boldsymbol{g}_{n'}\vert^2 P_{n'} + \sigma^2\right)^2}}{\left(\displaystyle\sum_{j\in \mathcal{N}\setminus \{n\}} B\alpha_{j}\frac{\vert\boldsymbol{\mu}_j^T \boldsymbol{g}_n\vert^2}{\ln (2) \left(\displaystyle\sum_{n' \in \mathcal{N}\setminus \{j\}} \vert\boldsymbol{\mu}_{j}^T \boldsymbol{g}_{n'}\vert^2 P_{n'} + \sigma^2\right)}+\lambda_n+\tilde{\eta}_{\textrm{Din}}^{(r-1)^*} + \nu_n \kappa A_{n,n}\right)^2} >0.
    \end{equation}
\end{table*}
\begin{table*}[t]
    \begin{align}
    \label{Scalability}
            &\theta I_n(\mathbf{p})-I_n(\theta\mathbf{p})\nonumber\\
            &=\theta \frac{B\alpha_{n}\log_{2}(\textrm{e})}{\displaystyle\sum_{j\in \mathcal{N}\setminus
            \{n \}}B\alpha_{j}\frac{\vert\boldsymbol{\mu}_j^T \boldsymbol{g}_n\vert^2}{ \left(\displaystyle\sum_{n'\in \mathcal{N}\setminus \{j \}}\vert\boldsymbol{\mu}_{j}^T \boldsymbol{g}_{n'}\vert^2 P_{n'} + \sigma^2\right)}+\lambda_n+\tilde{\eta}_{\textrm{Din}}^{(r-1)^*} + \nu_n \kappa A_{n,n}}
            -\frac{B\alpha_{n}\log_{2}(\textrm{e})}{\displaystyle\sum_{j\in \mathcal{N}\setminus \{n \}}B\alpha_{j}\frac{\vert\boldsymbol{\mu}_j^T \boldsymbol{g}_n\vert^2/\ln(2)}{\displaystyle\sum_{n'\in \mathcal{N}\setminus \{j \}}\vert\boldsymbol{\mu}_{j}^T \boldsymbol{g}_{n'}\vert^2 \theta P_{n'} + \sigma^2}+\lambda_n+\tilde{\eta}_{\textrm{Din}}^{(r-1)^*} + \nu_n \kappa A_{n,n}} \\
            &=\theta \frac{B\alpha_{n}\log_{2}(\textrm{e})}{\displaystyle\sum_{j\in \mathcal{N}\setminus \{n \}}B\alpha_{j}\frac{\vert\boldsymbol{\mu}_j^T \boldsymbol{g}_n\vert^2/\ln(2)}{\displaystyle\sum_{n'\in \mathcal{N}\setminus \{j \}}\vert\boldsymbol{\mu}_{j}^T \boldsymbol{g}_{n'}\vert^2 P_{n'} + \sigma^2}+\lambda_n+\tilde{\eta}_{\textrm{Din}}^{(r-1)^*} + \nu_n \kappa A_{n,n}}
            -\theta\frac{B\alpha_{n}\log_{2}(\textrm{e})}{\displaystyle\sum_{j\in \mathcal{N}\setminus \{n \}}B\alpha_{j}\frac{\vert\boldsymbol{\mu}_j^T \boldsymbol{g}_n\vert^2/ \ln (2)}{ \displaystyle\sum_{n'\in \mathcal{N}\setminus \{j \}}\vert\boldsymbol{\mu}_{j}^T \boldsymbol{g}_{n'}\vert^2  P_{n'} + \frac{\sigma^2}{\theta}}+\theta\lambda_n+\theta\tilde{\eta}_{\textrm{Din}}^{(r-1)^*} + \nu_n \theta \kappa A_{n,n}}. \nonumber
    \end{align} 
    \hrule
    \vspace*{-0.3cm}
\end{table*}
In \cite{Yates1995}, the definition of the standard interference function is introduced as below.
\begin{definition}\label{SIFdefinition}
The function $\boldsymbol{I}^{\normalfont{\textrm{ref}}}(\normalfont{\mathbf{p}}^{\normalfont{\textrm{ref}}})$ is a standard interference function if all of the following properties are satisfied with $\normalfont{\mathbf{p}}^{\normalfont{\textrm{ref}}} \succeq 0$. 
\begin{itemize}
\item Positivity: $\boldsymbol{I}^{\normalfont{\textrm{ref}}}(\normalfont{\mathbf{p}}^{\normalfont{\textrm{ref}}})>0$.
\item Monotonicity: If $\normalfont{\mathbf{p}}_{1}^{\normalfont{\textrm{ref}}}\succeq\normalfont{\mathbf{p}}_{2}^{\normalfont{\textrm{ref}}}$, then  $\boldsymbol{I}^{\normalfont{\textrm{ref}}}(\normalfont{\mathbf{p}}_{1}^{\normalfont{\textrm{ref}}})\succeq\boldsymbol{I}^{\normalfont{\textrm{ref}}}(\normalfont{\mathbf{p}}_{2}^{\normalfont{\textrm{ref}}})$.
\item Scalability: For any $\theta >1$, $\theta \boldsymbol{I}^{\normalfont{\textrm{ref}}}(\normalfont{\mathbf{p}}^{\normalfont{\textrm{ref}}})\succ\boldsymbol{I}^{\textrm{ref}}(\theta \normalfont{\mathbf{p}}^{\normalfont{\textrm{ref}}})$.
\end{itemize}
\end{definition}

To prove Lemma \ref{SIFlemma}, we will prove that the given $I_n(\mathbf{p})$ is indeed a standard interference function.
We will consecutively verify three properties of a standard interference function on the fixed-point equation in \eqref{PowerUpdate}. The positivity property of $I_n(\mathbf{p})$ is always guaranteed since both numerator and denominator are always positive. Next, we consider the first-order derivation of $I_n(\mathbf{p})$ with respect to $P_n$ as in \eqref{DerivativeIF}. Since  $I_n(\mathbf{p})$ is an increasing function due to $\frac{\partial I(\mathbf{p})}{\partial P_n}>0$, we have $I_n(\mathbf{p}_1) > I_n(\mathbf{p}_2)$ if $\mathbf{p}_1\succeq\mathbf{p}_2$. Therefore, the monotonicity property is satisfied. Finally, we verify the scalability property, for any real number $\theta>1$, the difference $\theta I_n(\mathbf{p})-I_n(\theta\boldsymbol{ P})$ is written as \eqref{Scalability} \textcolor{black}{at the top of next page}. We first observe that $\sum_{n'\in \mathcal{N}\setminus \{j\}}\vert\boldsymbol{\mu}_{j}^T \boldsymbol{g}_{n'}\vert^2  P_{n'} + \sigma^2/\theta<\sum_{n'\in \mathcal{N}\setminus \{j\}}\vert\boldsymbol{\mu}_{j}^T \boldsymbol{g}_{n'}\vert^2  P_{n'} + \sigma^2$ since $\theta > 1$, and thus the denominator of the first term is smaller than the denominator of the remaining term. Therefore, given any real number $\theta>1$, $\theta I_n(\mathbf{p})>I_n(\theta\boldsymbol{ P})$ is always true. Therefore, the scalability property is guaranteed.

The authors in \cite{Yates1995} then deduce the following Lemma.
\begin{lemma}\label{PowerControl}
If $\textbf{I}^{\normalfont{\textrm{ref}}}(\normalfont{\mathbf{p}})$ is a standard interference function and $\mathbf{p}^{\normalfont{\textrm{ref}}}\succeq \boldsymbol{I}^{\normalfont{\textrm{ref}}}(\mathbf{p}^{\normalfont{\textrm{ref}}})$ has a solution. Then, the following power allocation algorithm is provided
\begin{equation} \label{PowerControlAlgorithm}
\normalfont{\mathbf{p}}^{\textrm{ref},(m+1)}=\boldsymbol{I}^{\textrm{ref}}(\normalfont{\mathbf{p}}^{\textrm{ref},(m)}),
\end{equation}
where $m$ is the iteration index. \textcolor{black}{Starting from any initial power vector $\normalfont{\mathbf{p}}(0)$, the iterative power allocation in \eqref{PowerControlAlgorithm} always converges to a unique fixed-point.}
\end{lemma}
Thus the proof of Lemma \ref{SIFlemma} is completed.
\section{Proof of Lemma \ref{AlternativeLemma}} \label{ProofLemma4}
Let $\eta(\boldsymbol{\Psi}^{(t)},\boldsymbol{\tau}^{(t)},\mathbf{p}^{(t)})$ denotes the objective value after the $t$-th iteration in Algorithm \ref{AAlgorithm}. At the following $(t+1)$-th iteration, the solution from Algorithm \ref{Algorithm1} is the optimal user association to all UEs, and thus the value of objective function is non-decreasing $\eta(\boldsymbol{\Psi}^{(t)},\boldsymbol{\tau}^{(t)},\mathbf{p}^{(t)})\leq \eta(\boldsymbol{\Psi}^{(t+1)},\boldsymbol{\tau}^{(t)},\mathbf{p}^{(t)})$.
Next, after applying the Algorithm \ref{AlgorithmRRU}, we have the RRU activation $\boldsymbol{\tau}^{(t)}$ that achieves better overall EE, $\eta(\boldsymbol{\Psi}^{(t+1)},\boldsymbol{\tau}^{(t)},\mathbf{p}^{(t)})\leq \eta(\boldsymbol{\Psi}^{(t+1)},\boldsymbol{\tau}^{(t+1)},\mathbf{p}^{(t)})$. 
Then, given the information of the user association, from Lemma \ref{LogLemma}, applying the procedure as in Algorithm \ref{LogApproximationAlgorithm} always increases the overall EE, and thus we have ${\eta}(\boldsymbol{\Psi}^{(t+1)},\boldsymbol{\tau}^{(t+1)},\mathbf{p}^{(t)})\leq\eta(\boldsymbol{\Psi}^{(t+1)},\boldsymbol{\tau}^{(t+1)},\mathbf{p}^{(t+1)})$. We finally deduce the following result
\begin{equation}        
    \begin{aligned}
        \eta(\boldsymbol{\Psi}^{(t)},\boldsymbol{\tau}^{(t)},\mathbf{p}^{(t)}) 
        &\leq \eta(\boldsymbol{\Psi}^{(t+1)},\boldsymbol{\tau}^{(t)},\mathbf{p}^{(t)}) \\
        &\leq \eta(\boldsymbol{\Psi}^{(t+1)},\boldsymbol{\tau}^{(t+1)},\mathbf{p}^{(t)}) \\
        &\leq \eta(\boldsymbol{\Psi}^{(t+1)},\boldsymbol{\tau}^{(t+1)},\mathbf{p}^{(t+1)}).
    \end{aligned}                               
\end{equation} 
It means that the objective value of the problem in \eqref{Problemv1} is non-decreasing after each iteration of Algorithm \ref{AAlgorithm}. \textcolor{black}{Note that the denominator of the objective function does not vanish even when the transmit powers of all UEs are close to zero due to the existence of the circuit power and RRU consumed power}. Moreover, all variables of the problem in \eqref{Problemv1} are limited by constraints, and thus the EE value is bounded . Therefore, the convergence of Algorithm \ref{AAlgorithm} is guaranteed. 
Thus, this complete the proof of Lemma \ref{AlternativeLemma}.

\section{Proof of Proposition \ref{prop:DBCA-GNN}} \label{Proof:DBCA-GNN}

\begin{table*}[t]
    \begin{align}
        u_{n}^{(l)} &=  \Omega_{2}^{(l)} \left( u_{n}^{(l-1)}, \textrm{SUM} \left( \left\{ \Omega_{1}^{(l)} \left( v_{k}^{(l-1)} \right) \bigg\vert v_{k} \in \mathcal{A}(v_{n}^{\textrm{u}}) \right\} \right) \right) \quad, \mathcal{A}(u_{n}) = \{r_{k}, e_{kn} \vert k \in\mathcal{K}\} \label{proof:DBCA_GNN_first}\\
        r_{k}^{(l)} &=  \Omega_{4}^{(l)} \left( r_{k}^{(l-1)}, \textrm{SUM} \left( \left\{ \Omega_{3}^{(l)} \left(v_{n}^{(l-1)} \right) \bigg\vert v_{n} \in \mathcal{A}(r_{k}) \right\} \right) \right) \quad, \mathcal{A}(r_{k}) = \{u_{n}, e_{kn} \vert n \in\mathcal{N}\}\\
        e_{kn}^{(l)} &= \Omega_{5}^{(l)} \left( e_{kn}^{(l-1)}, \left\{ v \vert v \in \mathcal{A}(e_{kn})\right\}\right) \quad, \mathcal{A}(e_{kn}) = \{r_{k}^{(l)}, u_{n}^{(l)}  \}. \label{proof:DBCA_GNN_last}
    \end{align}
    \hrule
    \vspace*{-0.25cm}
\end{table*}

\textcolor{black}{The DBCA can be implemented in an agent-message passing fashion.} In particular, at each communication round $d$, every agent $i$ broadcasts a message based on its internal state $x^{(d-1)}_i$ to all neighboring agents,  formulated as 
\begin{align}
    m_{j,i}^{(d)} = \Xi_{1}^{(d)}\left(x^{(d-1)}_i\right) \quad, j\in \mathcal{A}(i), \forall i,
\end{align}
where $\Xi_{1}^{(d)}$ denotes the message generation function at round $d$.
Then, upon receiving the messages from its neighbors, each agent aggregates the incoming information and updates its internal state as follows
\begin{align}
    x^{(d)}_i = \Xi_{2}^{(d)}\left(x^{(d-1)}_i,  \Xi_{3}^{(d)}\left(\left\{m^{(l)}_{j,i}: j \in \mathcal{A}(i)\right\}\right)\right) \quad, \forall i.
\end{align}
where $\Xi_{2}^{(d)}$, and $\Xi_{3}^{(d)}$  represent the update and aggregation functions, respectively, in the current round.
This procedure is repeated for a fixed number of iterations or until convergence.

We denote the number of communication rounds in DBCA and the number of layers in ISAC-GNN as $S$. First, the update of DBCA at the $(d)$-th communication round is given as
\begin{align}
    \hat{m}_{j,i}^{(d)} &= \Xi_{1}^{(d)}\left(\hat{x}^{(d-1)}_i\right), \\
    \hat{x}^{(d)}_i &= \Xi_{2}^{(d)}\left(\hat{x}^{(d-1)}_i,  \Xi_{3}^{(d)}\left(\left\{\hat{m}^{(d)}_{j,i} \vert j \in \mathcal{A}(i)\right\}\right)\right) \quad, \forall i \nonumber
\end{align}
Next, we reformulate the node and edge update rule of the ISAC-GNN as shown in \eqref{proof:DBCA_GNN_first} through \eqref{proof:DBCA_GNN_last}. Then, we can obtain a general form for updating all elements in the $(l)$-th ISAC-GNN layer as
\begin{align}\label{proofeq:ISAC-GNN-modified}
    \begin{aligned}
        m(v_j)^{(l)} &=  \Psi_{1}^{(l)} \left( x_{j}^{(l-1)} \right), \\
        x_{v_i}^{(l)} &=  \Psi_{2}^{(l)} \left( x_{i}^{(l-1)}, \Psi_{3}\left\{ m(v_j)^{(l)} \vert v_j \in \mathcal{A}(v_i) \right\} \right).
    \end{aligned}
\end{align}
where $v_i\in \{ \mathbf{u} \cap \mathbf{r} \cap \mathbf{e} \}$, and $\mathbf{u} = \{u_{n} \vert n \in\mathcal{N} \}, \mathbf{r} = \{r_{k} \vert k \in\mathcal{K} \}, \mathbf{e} = \{e_{kn} \vert n \in\mathcal{N}, k \in\mathcal{K} \}$. $\Psi_{1} \in \{ \Omega_{1}, \Omega_{3} \}$, $\Psi_{2} \in \{ \Omega_{2}, \Omega_{4}, \Omega_{5} \}$, and $\Psi_{3} \in \{ \textrm{SUM}, \textrm{CONCAT} \}$.

\subsection{ISAC-GNN as a Generalization of DBCA}

First, we want to prove that, for every $0 \leq s \leq S$\ and $\{\Psi_{1}^{(l)}, \Psi_{2}^{(l)}, \Psi_{3}^{(l)} \}_{1 \leq l \leq s}$, there always exists $\{\Xi_{1}^{(d)}, \Xi_{2}^{(d)}, \Xi_{3}^{(d)} \}_{0\leq d \leq s}$ such that $\hat{x}_{i}^{(s)} = x_{v_i}^{(s)}$. We prove it by induction. Since the initial feature is the same for both models, the statement holds as $\hat{x}_{i}^{(0)} = x_{v_i}^{(0)}$. Then, assume the statement is true at $(s - 1)$-th iteration as $\hat{x}_{i}^{(s-1)} = x_{v_i}^{(s-1)}$, we can construct the update of DBCA at the $(s)$-th iteration as

\begin{align}
    \hat{x}^{(s)}_i &\overset{\phantom{(a)}}{=} \Xi_{2}^{(s)}\left(\hat{x}^{(s-1)}_i,  \Xi_{3}^{(s)}\left(\left\{\hat{m}^{(s)}_{j,i} \vert j \in \mathcal{A}(i)\right\}\right)\right) \quad,\forall i \\
     &\overset{(a)}{=} \Xi_{2}^{(s)}\left(x^{(s-1)}_i, \Xi_{3}^{(s)}\left(\left\{\Xi_{1}^{(s)}\left(x^{(s-1)}_i\right) \vert j \in \mathcal{A}(i)\right\}\right)\right), \nonumber
\end{align}
where $(a)$ is due to the assumption of the statement at the $(s-1)$-th iteration. Then, let $\Xi_{1}^{(s)} = \Psi_{1}^{(s)}, \Xi_{2}^{(s)} = \Psi_{1}^{(s)}, \Xi_{3}^{(s)} = \Psi_{1}^{(s)}$, we have
\begin{align}
    \hat{x}^{(s)}_i &\overset{\phantom{(b)}}{=} \Psi_{2}^{(s)}\left(x^{(s-1)}_i,  \Psi_{3}^{(s)}\left(\left\{\Psi_{1}^{(s)}\left(x^{(s-1)}_i\right) \vert j \in \mathcal{A}(i)\right\}\right)\right) \nonumber\\
    &\overset{(b)}{=} x^{(s)},
\end{align}
where $(b)$ is derived from the node update of ISAC-GNN at the $(s)$-th iteration as in \eqref{proofeq:ISAC-GNN-modified}.

\subsection{DBCA Reformulated as ISAC-GNN}
Next, we want to show that a DBCA can be viewed as an ISAC-GNN, i.e., for every $0 \leq s \leq S$\ and $\{\Xi_{1}^{(d)}, \Xi_{2}^{(d)}, \Xi_{3}^{(d)} \}_{1 \leq d \leq s}$, there always exists $\{\Psi_{1}^{(l)}, \Psi_{2}^{(l)}, \Psi_{3}^{(l)} \}_{0\leq l \leq s}$ such that $x_{v_i}^{(s)} = \hat{x}_{i}^{(s)}$. We also prove this by induction. Similar as the previous proof, the statement is true at the base of $x_{v_i}^{(0)} = \hat{x}_{i}^{(0)}$ since they have the same initial node feature. Assume the statement is true at $(s - 1)$-th iteration as $x_{v_i}^{(s-1)}=\hat{x}_{i}^{(s-1)}$, we can construct the update of ISAC-GNN at the $(s)$-th iteration as
\begin{align}\label{proofeq:ISAC-GNN-modified2}
    x_{v_i}^{(s)} &\overset{\phantom{(a)}}{=}  \Psi_{2}^{(s)} \left( x_{i}^{(s-1)}, \Psi_{3}\left\{ \Psi_{1}^{(s)} \left( x_{j}^{(s-1)} \right) \vert v_j \in \mathcal{A}(v_i) \right\} \right)  \nonumber\\
    &\overset{(a)}{=}  \Psi_{2}^{(s)} \left( \hat{x}_{i}^{(s-1)}, \Psi_{3}\left\{ \Psi_{1}^{(s)} \left( \hat{x}_{j}^{(s-1)} \right) \vert v_j \in \mathcal{A}(v_i) \right\} \right) \\
    &\overset{(b)}{=} \Xi_{2}^{(s)}\left(\hat{x}^{(s-1)}_i,  \Xi_{3}^{(s)}\left(\left\{\Xi_{1}^{(s)}\left(\hat{x}^{(s-1)}_i\right) \vert j \in \mathcal{A}(i)\right\}\right)\right)  ,\nonumber
\end{align}
where $(a)$ is due to the assumption of the statement at the $(s-1)$-th iteration, $(b)$ is derived if we let $\Psi_{1}^{(s)} = \Xi_{1}^{(s)} , \Psi_{1}^{(s)} = \Xi_{2}^{(s)} , \Psi_{1}^{(s)} =\Xi_{3}^{(s)}$. Then, $x_{v_i}^{(s)} = \hat{x}_{i}^{(s)}$.
The proof is completed.

\section{Proof of Lemma \ref{lemma:DBCA-AA}} \label{Proof:DBCA-AA}

We model the considered system in a similar manner to graph modeling in Section \ref{sec:graph_model}. At the $(t)$-th iteration of the Algorithm~\ref{AAlgorithm}, the internal state for the $n$-th UE-agent is $s_{n}^{\textrm{u},(t)}=[P^{\textrm{max}}_{n}, P_{\textrm{cir}},\kappa,\hat{x}_{n}, \hat{y}_{n}, P_{n}^{(t)}]$, the internal state for the $k$-th RRU-agent is $s_{k}^{\textrm{r},(t)}=[\tilde{x}_{k}, \tilde{y}_{k}, P_{\textrm{sens}}, \tau_{k}^{(t)}]$, and the internal state for the edge-agent between the $n$-th UE and the $k$ RRU is given as $s_{k,n}^{\textrm{e},(t)}=[g_{kn}, \mu_{kn}^{(t)}]$. 
In this model, each RRU-agent is connected to all UE-agents, whereas each edge-agent connects only to its corresponding UE-agent and RRU-agent, i.e, the $(k,n)$ edge-agent only connects to the $n$-th UE and the $k$-th RRU-agent. Then, we can rewrite the alternative optimization algorithm in Algorithm~\ref{AAlgorithm} as an agent-based iterative procedure as in Algorithm~\ref{alg:iterative_optimization}, where the $\mathcal{U}^{(t)}$, $\mathcal{T}^{(t)}$, $\mathcal{P}^{(t)}$ are the solver functions for the user association problem in \eqref{ProblemAP1}, the RRU activation problem in \eqref{ProblemRRUV2}, and the power allocation problem in \eqref{ProblemP0} at the $(t)$-th iteration, respectively.

\begin{algorithm}[t]
\caption{Agent-based Iterative Alternative Optimization}
\label{alg:iterative_optimization}
\begin{algorithmic}[1]
\footnotesize
\STATE \textbf{Initialize} $s_{n}^{\textrm{u},(0)}, s_{k}^{\textrm{r},(t)}, s_{kn}^{\textrm{e},(0)}, \forall n\in\mathcal{N}, k \in\mathcal{K}.$
\FOR{iteration $t = 1, 2, \dots$}
    \STATE edge-agent $(k,n)$ update its internal state \\
    \vspace{0.5em}
    $\mu_{kn}^{(t)} = \mathcal{U}^{(t)}\left(s_{kn}^{\textrm{e},(t-1)}, \left\{s_{n}^{\textrm{u},(t-1)}, s_{k}^{\textrm{r}} \right\} \right).$
    \vspace{0.5em}
    \STATE \textcolor{black}{RRU-agent $k$ updates its internal state} \\
    \vspace{0.5em}
    $\tau_{k}^{(t)} = \mathcal{T}^{(t)}\left( s_{k}^{\textrm{r},(t-1)}, \left\{s_{n}^{\textrm{u}}, s_{kn}^{\textrm{e},(t-1)} \vert \forall n \in\mathcal{N} \right\} \right).$
    \STATE \textcolor{black}{UE-agent $n$ updates its internal state} \\
    \vspace{0.5em}
    $P_{n}^{(t)} = \mathcal{P}^{(t)}\left( s_{n}^{\textrm{u},(t-1)}, \left\{s_{k}^{\textrm{r}}, s_{kn}^{\textrm{e},(t-1)} \vert \forall k \in\mathcal{K} \right\} \right).$
\ENDFOR
\end{algorithmic}
\end{algorithm}

Now, we want to show that the alternative optimization algorithm with $T$ iterations is a DBCA with at most $3T$ iterations. Particularly, at the $(3t)$ iteration, all the edge-agents update their internal state, while the UE-agents update their internal state at the $(3t + 1)$ iteration, and the RRU-agent update their internal state at the $(3t + 2)$ iteration. In detailed, at the $(3t)$-th iteration, all RRU-agents, and UE-agents broadcast their internal state to their neighboring edge-agents as below
\begin{align}
    m_{n}^{\textrm{u,e},(3t)} &= \Xi_{1}^{(3t)}\left(s_{n}^{\textrm{u},(3t-1)}\right) \\
    &= s_{n}^{\textrm{u},(3t-1)}= [P^{\textrm{max}}_{n}, P_{\textrm{cir}},\kappa,\hat{x}_{n}, \hat{y}_{n}, P_{n}^{(3t-1)}], \nonumber\\
    m_{k}^{\textrm{r,e},(3t)} &= \Xi_{1}^{(t)}\left(s_{k}^{\textrm{r},(3t-1)}\right) \\&= s_{k}^{\textrm{r},(3t-1)} =[\tilde{x}_{k}, \tilde{y}_{k}, P_{\textrm{sens}}, \tau_{k}^{(3t-1)}].\nonumber
\end{align}
Edge-agent receives all incoming messages and updates its internal state as
\begin{align}
    \mu_{kn}^{(3t)}  &= \Xi_{2}^{(3t)}\left(s_{k,n}^{\textrm{e},(3t-1)}, \Xi_{3}^{(3t)}\left(\left\{m_{n}^{\textrm{u,e},(3t)}, m_{k}^{\textrm{r,e},(3t)} \right\}\right)\right) \nonumber\\
    &= \mathcal{U}^{(3t)}\left(s_{k,n}^{\textrm{e},(3t-1)}, \left\{s_{n}^{\textrm{u},(3t-1)}, s_{k}^{\textrm{r},(3t-1)} \right\} \right),
\end{align}
which is recognized as the agent-update in the DBCA at the $(3t)$ iteration. 
Next, we investigate the update at the $(3t+1)$-th iteration. In the same manner as the previous iteration, all edge-agents and UE-agents broadcast their internal states to their neighboring RRU-agents as below
\begin{align}
    m_{k,n}^{\textrm{e,u},(3t+1)} &= \Xi_{1}^{(3t+1)}\left(s_{k,n}^{\textrm{e},(3t)}\right) = s_{k,n}^{\textrm{e},(3t)} =[g_{kn}, \mu_{kn}^{(3t)}], \\
    m_{k}^{\textrm{r,u},(3t+1)} &= \Xi_{1}^{(3t+1)}\left(s_{k}^{\textrm{r},(3t)}\right) \\
    &= s_{k}^{\textrm{r},(3t)} =[\tilde{x}_{k}, \tilde{y}_{k}, P_{\textrm{sens}}, \tau_{k}^{(3t)}]. \nonumber
\end{align}
RRU-agent receives all incoming messages and updates its internal state as
\begin{align}
    \tau_{k}^{(3t+1)}  &= \Xi_{2}^{(3t+1)}\left(s_{k,n}^{\textrm{e},(3t)}, \right. \nonumber \left.  \Xi_{3}^{(3t+1)}\left(\left\{m_{k,n}^{\textrm{e,u},(3t)}, m_{k}^{\textrm{r,u},(3t+1)} \vert \forall k \right\}\right)\right)\nonumber\\
    &= \mathcal{T}^{(3t+1)}\left(s_{n}^{\textrm{u},(3t)}, \left\{s_{k,n}^{\textrm{e},(3t)}, s_{k}^{\textrm{r},(3t)} \vert \forall k \right\} \right) ,
\end{align}
which is recognized as the agent-update in the DBCA at the \mbox{$(3t+1)$}-th iteration. 
Then at the $(3t+2)$-th iteration, all edge-agents and RRU-agents broadcast their internal states to their neighboring UE-agents as below
\begin{align}
    m_{k,n}^{\textrm{e,u},(3t+2)} &= \Xi_{1}^{(3t+2)}\left(s_{k,n}^{\textrm{e},(3t+1)}\right) = s_{k,n}^{\textrm{e},(3t+1)} =[g_{kn}, \mu_{kn}^{(3t+1)}], \\
    m_{k}^{\textrm{r,u},(3t+2)} &= \Xi_{1}^{(3t+2)}\left(s_{k}^{\textrm{r},(3t+1)}\right)  \\
    &= s_{k}^{\textrm{r},(3t+1)}=[\tilde{x}_{k}, \tilde{y}_{k}, P_{\textrm{sens}}, \tau_{k}^{(3t+1)}]. \nonumber
\end{align}
UE-agent receives all incoming messages and updates its internal state as
\begin{align}
    P_{n}^{(3t+2)} &= \Xi_{2}^{(3t+2)}\left(s_{k,n}^{\textrm{e},(3t+1)}, \right.\nonumber \\
    & \left.\hspace{4em} \Xi_{3}^{(3t+2)}\left(\left\{m_{k,n}^{\textrm{e,u},(3t+2)}, m_{k}^{\textrm{r,u},(3t+2)} \vert \forall k \right\}\right)\right) \nonumber \\
    &= \mathcal{P}^{(3t+2)}\left(s_{n}^{\textrm{u},(3t+1)}, \left\{s_{k,n}^{\textrm{e},(3t+1)}, s_{k}^{\textrm{r}(3t+1)} \vert \forall k \right\} \right),
\end{align}
which is recognized as the agent-update in the DBCA at the $(3t+2)$ iteration. 
Thus, this complete the proof of Lemma \ref{lemma:DBCA-AA}.



\bibliographystyle{IEEEtran}
\balance
\bibliography{reference}








\end{document}